\documentclass[12pt]{article}

\usepackage{epsfig}
\usepackage{latexsym}
\usepackage{axodraw}
\input{amssymb.sty}

\def\bbz{{\mathbb Z}}

\def\frac#1#2{{{#1}\over{#2}}}
\def\tfrac#1#2{{\textstyle{{#1}\over{#2}}}}

\def\half{\tfrac{1}{2}}

\def\quart{\tfrac{1}{4}}

\newcommand{\fdown}{{\mbox{$F_{\mu\nu}$}}}
\newcommand{\fup}{{\mbox{$F^{\mu\nu}$}}}

\newcommand{\fstarup}{{\mbox{$\mbox{}^*\!F^{\mu\nu}$}}}

\newcommand{\epsup}{{\mbox{$\epsilon^{\mu\nu\rho\sigma}$}}}

\newcommand{\bq}{\begin{equation}}
\newcommand{\dq}{\end{equation}}

\begin{document}

\begin{titlepage}

\baselineskip 24pt

\begin{center}

{\Large {\bf Fermion Generations and Mixing from Dualized Standard Model}}
\footnote{Course of lectures given at the 42nd Cracow School of
Theoretical Physics, 2002.}

\vspace{.5cm}

\baselineskip 14pt

{\large CHAN Hong-Mo}\\
h.m.chan\,@\,rl.ac.uk \\
{\it Rutherford Appleton Laboratory,\\
  Chilton, Didcot, Oxon, OX11 0QX, United Kingdom}\\
\vspace{.5cm}
{\large TSOU Sheung Tsun}\\
tsou\,@\,maths.ox.ac.uk\\
{\it Mathematical Institute, University of Oxford,\\
  24-29 St. Giles', Oxford, OX1 3LB, United Kingdom}

\end{center}

\vspace{.3cm}

\begin{abstract}

The puzzle of fermion generations is generally recognized as one of the
most outstanding problems of present particle physics.  In these lectures,
we review a possible solution based on a nonabelian generalization of
electric--magnetic duality derived some years ago.  This nonabelian
duality implies the existence of another $SU(3)$ symmetry dual to colour,
which is necessarily broken when colour is confined and so can play the
role of the ``horizontal'' symmetry for fermion generations.  When thus
identified, dual colour then predicts 3 and only 3 fermion generations,
besides suggesting a special Higgs mechanism for breaking the generation
symmetry.  A phenomenological model with a Higgs potential and a Yukawa
coupling constructed on these premises is shown to explain immediately
all the salient qualitative features of the fermion mass hierarchy and
mixing pattern, excepting for the moment CP-violation.  In particular,
though treated on exactly the same footing, quarks and leptons are seen
to have very different mixing patterns as experimentally observed, with
leptons having generally larger mixings than quarks.  The model offers
further a perturbative method for calculating mixing parameters and mass
ratios between generations.  Calculations already carried out to 1-loop
order is shown to give with only 3 adjustable parameters the following
quantities all to within present experimental error: all 9 CKM matrix
elements $|V_{rs}|$ for quarks, the neutrino oscillation angles or the
MNS lepton mixing matrix elements $|U_{\mu 3}|, |U_{e 3}|$, and the mass
ratios $m_c/m_t, m_s/m_b, m_\mu/m_\tau$.  The special feature of this
model crucial for deriving the above results is a fermion mass matrix
which changes its orientation (rotates) in generation space with changing
energy scale, a feature which is shown to have direct empirical support,
and although potentially dangerous for flavour-violation is found through
detailed analysis not to be the case.  With its parameters now so fitted,
the resulting scheme is highly predictive giving in particular correlated
predictions in low energy FCNC effects (meson mass splittings and decays,
$\mu-e$ conversion in nuclei, etc.) and in ultra-high energy (post-GZK)
air showers from cosmic rays, both of which can hopefully be tested soon
by experiment.

\end{abstract}

\end{titlepage}

\clearpage

\baselineskip 14pt

\setcounter{equation}{0}

\section{Introduction}

As far as we know today, quarks and leptons, the fermionic fundamental
building blocks of our material world, each occurs in 3, and apparently
only 3, copies called generations having very similar properties apart
from their masses.  The masses, however, vary greatly, dropping from
generation to generation by about one to more than two orders of magnitude
depending on the fermion species.  For charged leptons and quarks, the
masses are now quite well determined and are listed in the Particle
Physics Booklet \cite{databook} as follows:
\begin{equation}
\begin{array}{lll}
m_t \sim 175\ {\rm GeV}, & m_c \sim 1.2\ {\rm GeV},
 &  m_u \sim 3\ {\rm MeV}; \\
m_b  \sim  4.2\ {\rm GeV}, & m_s \sim 120\ {\rm MeV},
 &  m_d \sim 6\ {\rm MeV};  \\
m_\tau  = 1.777\ {\rm GeV}, & m_\mu = 105.6\ {\rm MeV},
 &  m_e = 0.51\ {\rm MeV}.
\label{masses}
\end{array}
\end{equation}
For neutrinos, the picture is not yet as clear, but with the recent
discovery of $\nu_\tau$, and observation of neutrino
oscillations with measurement of some of the relevant parameters,
a similar pattern looks increasingly likely
to emerge, namely again 3 generations of neutrinos with a hierarchical
mass spectrum.

That this should be the case has long been regarded theoretically as
quite a mystery.  First of all, that nature should want several species
of fermions with different quantum numbers and interactions to  build
her multifarious world seems understandable, but why 3 copies of each?
And, what is more, why has she given them so different masses?  Indeed,
the general theoretical idea is that particles get their masses mostly
from self-energy through their interactions.  Why then these widely
different masses for the 3 generations which have as far as we know
identical interactions?  In fact, long before the full picture is known,
the existence of the muon has already prompted Feynman to post above
his bed the famous question: ``Why does the muon weigh?''  And now, with
3 generations in each of all 4 fermion species, and each generation
weighing more than the next by large factors, Feynman's question has
become even more pressing.

And the mystery does not end there.  With more empirical information
accumulated, another puzzling phenomenon soon revealed itself.  The
12 fermion states of different generations and species can each be
represented by a state vector in 3-dimensional generation space.
Within each species, the 3 generations are independent quantum states
and should thus be represented by orthogonal vectors forming together
an orthonormal triad.  For quarks, for example, the 3 up quark states
$t, c, u$ form together a $U$ triad, while the 3 down quark states
$b, s, d$ form together a $D$ triad.  The question, first posed by
Cabbibo \cite{Cabbibo}, then arises, namely whether the $U$ and $D$
triads are the same, and if not, how they are related.  Now
the relative orientations, namely the inner (or dot) products, between
any pairs of vectors in the 2 triads can be inferred empirically from
experiment on e.g.\ hadron decays.  The matrix of these inner products
is then the famous CKM matrix \cite{Cabbibo,CKM}, for which the latest
empirical information is summarized in \cite{databook} as follows:
\begin{eqnarray}
\lefteqn{
\left( \begin{array}{ccc} |V_{ud}| & |V_{us}| & |V_{ub}| \\
                          |V_{cd}| & |V_{cs}| & |V_{cb}| \\
                          |V_{td}| & |V_{ts}| & |V_{tb}|
   \end{array} \right)
   =} \nonumber\\
&& \left( \begin{array}{lll}
   0.9742-0.9757 & 0.219-0.226 & 0.002-0.005 \\
   0.219-0.225 & 0.9734-0.9749 & 0.037-0.043 \\
   0.004-0.014 & 0.035-0.043 & 0.9990-0.9993  \end{array} \right).
\label{exckm}
\end{eqnarray}
One notices that the $U$ and $D$ triads are indeed not aligned but are
nevertheless tantalisingly close to being so, namely that the CKM matrix
is close to being the unit matrix.  One notices also that the off-diagonal
(mixing) elements seem to have hierarchical values with $|V_{us}|, |V_{cd}|
\gg |V_{cb}|, |V_{ts}| \gg |V_{ub}|, |V_{ts}|$.

The same question can be repeated for leptons, i.e.\ on the relative
orientation between the up and down triads, namely between the triad $L$
of the charged leptons $\tau, \mu, e$ and the triad $N$ of the neutrino
mass eigenstates traditionally denoted in order of decreasing mass
by $\nu_3, \nu_2, \nu_1$.  The matrix of inner products between pairs
of vectors, one from each triad, is in this case known at the MNS
matrix \cite{MNS}, the elements of which are measured in neutrino
oscillation experiments.
So far, experiments on atmospheric neutrinos from $\mu$
decay \cite{SuperK,Soudan} have shown that the mixing between the muon
neutrino and the heaviest mass eigenstate $\nu_3$, namely the MNS element
$U_{\mu3}$, is near maximal.  Those on solar neutrinos \cite{SuperK,Sno,
Homestake,Sage,Gallex} measure the mixing between the electron neutrino
and the second heaviest mass eigenstate $\nu_2$, namely the MNS element
$U_{e2}$, while reactor experiments such as CHOOZ \cite{Chooz} have
given bounds on the mixing between the electron neutrino and $\nu_3$,
namely the MNS element $U_{e3}$.  The total empirical information on the
MNS matrix available to-date is briefly summarized below:
\begin{equation}
\left( \begin{array}{ccc} |U_{e1}| & |U_{e2}| & |U_{e3}| \\
                          |U_{\mu1}| & |U_{\mu2}| & |U_{\mu3}| \\
                          |U_{\tau1}| & |U_{\tau2}| & |U_{\tau3}|
   \end{array} \right)
   = \left( \begin{array}{ccc} \star & 0.4-0.7 & 0.0-0.15 \\
                               \star & \star & 0.56-0.83 \\
                               \star & \star & \star \end{array} \right).
\label{exmns}
\end{equation}
There are actually several solutions to the solar neutrino problem still
consistent with present experiment, among which the so-called large mixing
angle MSW \cite{MSW} solution is the most favoured and is the one quoted
in (\ref{exmns}).  One notices that in contrast to the CKM matrix, the
MNS matrix is far from diagonal, with some off-diagonal elements very large,
but still the corner element $U_{e3}$ is much smaller than the other two.

Thus, together with the markedly hierarchical mass spectra, the mixing
patterns of quarks and leptons constitute a vast amount of quantitative
data needing theoretical understanding.  In spite of its many successes,
however, the Standard Model as conventionally formulated offers no
explanation at all either for the existence of the 3 fermion generations
in the first place, nor yet for their striking mass and mixing patterns,
but takes instead all these features just as fundamental inputs.  Indeed,
fermion masses and mixings together account for some three quarters of
the twenty odd parameters defining the Standard Model, which would be
dramatically reduced if some understanding of the generation puzzle can
somehow be achieved.  For this reason, the solution of the generation
puzzle is justly regarded by many as one of the most urgent problems
facing particle physics today.

In these lectures, we wish to describe a possible solution to the problem
based on a nonabelian generalization of electric--magnetic duality.  It is
a solution within the Standard Model framework, without introducing, for
example, either supersymmetry or higher dimensions, although it is not,
as far as is known, inconsistent with either of these extensions.
Apart from
offering right from the start a {\it raison d'\^etre} for 3 generations of
fermions, this scheme, which we call the Dualized Standard Model (DSM),
explains the fermion mass hierarchy and the mixing phenomena and suggests
even a perturbative method for calculating mass and mixing parameters.
Calculations with it have been carried out so far to the 1-loop level, and
the score to-date is as follows.  With 3 real parameters fitted to data,
it gives correctly to within present experimental bounds the following
measured quantities: the mass ratios $m_c/m_t, m_s/m_b, m_\mu/m_\tau$, all
9 elements $|V_{rs}|$ of the CKM matrix, plus  the 2 elements $|U_{\mu3}|$ and $|U_{e3}|$ of the MNS matrix measured in neutrino oscillation experiments.
It gives further by interpolation sensible though inaccurate estimates for
the following quantities which are formally beyond the scope of the 1-loop
calculation so far performed: the mass ratios $m_u/m_t, m_d/m_b, m_e/m_\tau$
and the solar neutrino angle $U_{e2}$. These calculated and estimated
quantities represent altogether 12 independent fundamental parameters of
the Standard Model, which are thereby replaced by only 3 fitted parameters
in the DSM.  Next, with nearly all its parameters now fixed, the scheme
becomes highly predictive.  In particular, numerous detailed predictions
have been made in flavour-violation effects over a wide area comprising
meson mass differences, rare hadron decays, $e^+ e^-$ collisions, and
muon--electron conversion in nuclei.  Further predictions have been made
on effects as far apart in energy as neutrinoless double-beta decays in
nuclei and cosmic ray air showers beyond the GZK cut-off of $10^{20}$ eV
at the extreme end of the present observable energy range.  Wherever
possible, these predictions have been confronted with data, and so far,
all are found to remain within present empirical bounds, although a few
of them so closely as should be accessible soon to new experimental
tests.

Of course, that the DSM scheme seems to have largely succeeded in its
primary aim of explaining fermion generations and their mass and mixing
patterns, and at the same time to have survived all other tests to-date,
still does not mean that its tenets are thereby proved correct.  Stress
should thus be given to examining the result to see which of its basic
assumptions are really essential for obtaining the claimed agreement with
experiment.  At the same time, attention has to be paid to any aspects
in the scheme which can potentially be improved.  We hope to cover most
of these topics, though some only briefly, in the course of these lectures.

\setcounter{equation}{0}

\section{Electric--magnetic duality and its nonabelian generalization}

Let us start, however, from the beginning, with a reminder of ordinary
electric--magnetic duality and a review of its extension to nonabelian
Yang--Mills theory, then see eventually how it leads one to consider the
Dualized Standard Model for an explanation of fermion generations.  For
the reader interested mainly in the phenomenological aspects of DSM and
not so much in its theoretical basis, only a cursory look at this section
is needed, since no mastery of the details contained in here is required
for appreciation of the material in the later sections.

The Maxwell equations for electromagnetism are usually written as:
\begin{equation}
\begin{array}{ll} \left.
\begin{array}{r} {\rm div}\ \mathbf{E} = \rho \\ {\rm curl}\ \mathbf{B} -
\partial \mathbf{E}/\partial t = \mathbf{J} \end{array} \right\}
 & \partial_\nu F^{\mu\nu} =-j^\mu \\
& \\
\left.
\begin{array}{r} {\rm div}\
\mathbf{B} =0 \\ {\rm curl}\ \mathbf{E} + \partial \mathbf{B}/\partial
t =0 \end{array} \right\} & \partial_\nu \fstarup =0,
\end{array}
\end{equation}
where the {\em dual field tensor} \ \fstarup\ is defined to be
\bq
\fstarup = -\half \epsilon^{\mu\nu\rho\sigma}
F_{\rho\sigma}.  \label{hodge}
\dq

We see immediately that in the absence of matter, classical Maxwell
theory is invariant under duality:
\begin{eqnarray}
\partial_\nu {}^*\!F^{\mu\nu}&=&0 \quad [{\rm d}\,F=0]
\label{freemax1}
\\
\partial_\nu F^{\mu\nu}&=&0 \quad [{\rm d}\, {}^*\!F =0]
\label{freemax2}
\end{eqnarray}
where in square brackets are displayed the equivalent equations in the
language of differential forms.   Then by the Poincar\'e lemma we
deduce directly the existence of potentials $A$ and $\tilde{A}$
such that
\begin{eqnarray}
F_{\mu\nu}(x)& =& \partial_\nu A_\mu(x) - \partial_\mu A_\nu(x)\quad
[\,F={\rm d}A\,],
\label{fcurl}\\
{}^*\!F_{\mu\nu}(x)& =& \partial_\nu \tilde{A}_\mu(x) - \partial_\mu
\tilde{A}_\nu(x)\quad [\,{}^*F={\rm d}{\tilde{A}}\,]. \label{fstarcurl}
\end{eqnarray}
The two potentials transform independently under independent gauge
transformations $\Lambda$ and $\tilde{\Lambda}$:
\begin{eqnarray}
A_\mu(x) &\mapsto& A_\mu(x) + \partial_\mu \Lambda(x),\\
\tilde{A}_\mu(x) & \mapsto & \tilde{A}_\mu(x) + \partial_\mu
   \tilde{\Lambda}(x),
\end{eqnarray}
which means that the full symmetry of this theory is actually
$U(1) \times \tilde{U}(1)$, where the tilde on the second $U(1)$
indicates it is the symmetry of the dual potential $\tilde{A}$.  It is
important to note that the physical degrees of freedom remain just
either $F$ or ${}^*\!F$, not both, since $F$ and ${}^*\!F$ are related
by an {\em algebraic} equation (\ref{hodge}).  The dual symmetry is
there all the time but just physically not so readily detected and it
means that what we call `electric' or `magnetic' is entirely a matter
of choice.

Before we go back to discuss matter carrying charges of the gauge
theory, let us first distinguish between two types of charges:
sources and monopoles.  These are defined with respect to the gauge
field, which in turn is derivable from the gauge potential.

{\em Source charges}
are those charges that give rise to a
nonvanishing divergence of the field.  For example, the electric
current $j$ due to the presence of the electric charge $e$ occurs on
the right hand side of the first Maxwell equation, and is given in
the quantum case by
\begin{equation}
j^\mu= e\, \bar{\psi} \gamma^\mu \psi.
\end{equation}
In the Yang--Mills case with general nonabelian gauge group $G$, 
the first Maxwell equation is replaced by
the Yang--Mills equation:
\begin{equation}
 D_\nu F^{\mu\nu} = -j^\mu, \quad j^\mu= g\, \bar{\psi} \gamma^\mu
 \psi,
\end{equation}
where we define the covariant derivative $D$ by
\begin{equation}
D_\mu F^{\mu\nu}= \partial_\mu F^{\mu\nu}
-ig\,[\,A_\mu,F^{\mu\nu}\,].
\end{equation}

\begin{figure}[ht]
\vspace*{10mm}
\centering
\input{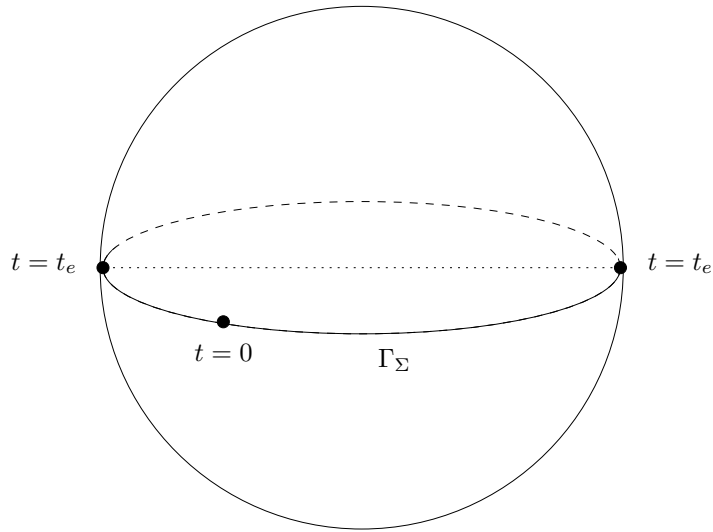}
\caption{An $SO(3)$ monopole.}
\label{so3mono}
\end{figure}

{\em Monopole charges}, on the other hand, are topological obstructions
specified geometrically by nontrivial $G$-bundles over every 2-sphere
$S^2$ surrounding the charge\footnote{For the nonmathematical reader,
a more intuitive picture of a monopole as topological obstruction can
be found in, for example, \cite{ourbook}, section 2.1.}.
They are classified by elements of
$\pi_1(G)$, the fundamental group of $G$, that is,
classes of closed loops in the group manifold which
can be continuously deformed into one another.
They are
typified by the (abelian)
magnetic monopole as first discussed by Dirac in 1931 \cite{Dirac}.   
A nonabelian example is that of $SO(3)$, where the monopole
charges are just denoted by a sign: $\pm 1$, with $+1$ corresponding
to the vacuum and $-1$ to the monopole.  Figure~\ref{so3mono}
illustrates this case.
Moreover, we can obtain the {\em Dirac
quantization condition} quite easily from the definition of the
monopole, which in the abelian case is:
\bq
e \tilde{e} = 2\pi,
\dq
and in the nonabelian case is:
\bq
g \tilde{g} =4\pi,
\label{Diraccond}
\dq
the difference between the two cases being only a matter of
conventional normalization \cite{ourbook,dualcomm}.

Now in the presence of electric charges, the Maxwell equations appear
usually as
\begin{eqnarray}
\partial_\nu {}^*\!F^{\mu\nu}&=&0 \label{max1}\\
\partial_\nu F^{\mu\nu}&=& -j^\mu.
  \label{max2}
\end{eqnarray}
The apparent asymmetry in these equations comes from the experimental
fact that there is only one type of charges
observed in nature which we choose
to regard as a source of the field $F$ (or, equivalently but
unconventionally, as a monopole of the field ${}^*\!F$).
But as we see by dualizing equations (\ref{max1}) and (\ref{max2}),
that is, by interchanging the role of electricity and magnetism in
relation to $F$,
we could
equally have thought of
these instead as source charges of the field ${}^*\!F$ (or,
similarly to the above, as monopoles of $F$):
\begin{eqnarray}
\partial_\nu {}^*\!F^{\mu\nu}&=& -\tilde{\jmath}^\mu \label{maxdual1}\\
\partial_\nu F^{\mu\nu}&=& 0.
 \label{maxdual2}
\end{eqnarray}
And if both electric and magnetic charges existed in nature,
then we would have the
dual symmetric pair:
\begin{eqnarray}
 \partial_\nu {}^*\!F^{\mu\nu}&=& -\tilde{\jmath}^\mu \\
\partial_\nu F^{\mu\nu}&=& -j^\mu.
\end{eqnarray}

The duality in the presence of matter goes in fact much deeper, as
can be seen if we use the Wu--Yang criterion \cite{WuYang,Chanstsou2}
to derive the Maxwell equations\footnote{What we present here is not
the textbook derivation of Maxwell's equations from an action, but
we consider this method to be much more intrinsic and geometric.}.
Consider first pure electromagnetism.  The free Maxwell action is:
\begin{equation}
{\mathcal A}_F^0 = - \quart \int \fdown \fup.
\label{actionff}
\end{equation}
The true variables of the (quantum) theory are the $A_\mu$, so
in (\ref{actionff}) we should put in a constraint to say that $\fdown$
is the curl of $A_\mu$ (\ref{fcurl}).  This can be viewed as a
topological constraint, because it is precisely equivalent to
(\ref{freemax1}).   Using the method of Lagrange multipliers, we form
the constrained action
\begin{equation}
{\mathcal A} = {\mathcal A}^0_F + \int \lambda_\mu
\,(\partial_\nu \fstarup)\,,
\label{actioncon}
\end{equation}
which we can now vary with respect to \fdown, obtaining
\bq
\fup = 2 \,\epsup \,\partial_\rho \lambda_\sigma \label{lagrange}
\dq
which implies (\ref{freemax2}).   Moreover, the Lagrange multiplier
$\lambda$ is exactly the dual potential $\tilde{A}$.
The derivation is entirely dual symmetric, since we can equally well
use (\ref{freemax2}) as constraint for the action ${\mathcal A}_F^0$,
now considered as a functional of \fstarup:
\bq
{\mathcal A}^0_F = \quart
\int {}^*\!F_{\mu\nu} {}^*\!F^{\mu\nu},
\dq
and obtain (\ref{freemax1}) as the equation of motion.

This method applies to the interaction of charges and fields as well.
In this case we start with the free field plus free particle action:
\bq
{\mathcal A}^0 = {\mathcal A}^0_F + \int  \bar{\psi} \,(i
\partial_\mu \gamma^\mu -m) \,\psi,
\dq
where we assume the free particle $m$ to satisfy the Dirac equation.
To fix ideas, let us regard this particle carrying an electric charge
$e$ as a monopole of the potential $\tilde{A}_\mu$.
Then the constraint we put in is (\ref{max2}):
\bq
{\mathcal A}' = {\mathcal A}^0 + \int \tilde{\lambda}_\mu \,(\partial_\nu
\fup + \jmath^\mu)\,.
\label{diracdual}
\dq
Varying with respect to ${}^*\!F$ gives us (\ref{max1}), and varying
with respect to $\bar{\psi}$ gives
\bq
(i \,\partial_\mu \gamma^\mu -m) \,\psi = - e A_\mu
\gamma^\mu \psi.
\label{diraceom}
\dq
So the complete set of equations for a Dirac particle carrying an
electric charge $e$
in an
electromagnetic field is (\ref{max1}), (\ref{max2}) and
(\ref{diraceom}).  The duals of
these equations will describe the dynamics of a Dirac magnetic
monopole in an electromagnetic field.

We see from this that the Wu--Yang criterion actually gives us an
intuitively clear picture of interactions.  The assertion
that there is a monopole at a certain spacetime point $x$ means that
the gauge field on a 2-sphere surrounding $x$ has to have a certain
topological configuration (e.g.\ giving a nontrivial bundle of a
particular class), and if the monopole moves to another point, then
the gauge field will have to rearrange itself so as to maintain the
same topological configuration around the new point.  There is thus
naturally a coupling between the gauge field and the position of the
monopole, or in
physical language a topologically induced interaction between the field
and the charge~\cite{WuYang}.  Furthermore, this treatment of
interaction between field and matter is entirely dual symmetric.

The next natural step is to generalize this duality
to the nonabelian Yang--Mills case.  Although there is no difficulty
in defining \fstarup, which is again given by (\ref{hodge}), we
immediately come to difficulties in the relation between field and
potential:
\bq
F_{\mu\nu} (x)=\partial_\nu A_\mu (x) - \partial_\mu A_\nu (x) + ig
\,[\,A_\mu (x), A_\nu (x)\,].
\label{ymfcurl}
\dq
First of all, despite appearances the Yang--Mills equation (in the
free field case)
\bq
D_\nu F^{\mu\nu} =0 \label{freeym}
\dq
and the Bianchi identity
\bq
D_\nu \fstarup =0 \label{bianchi}
\dq
are not dual-symmetric,  because the correct dual of the Yang--Mills
equation ought to be
\bq
\tilde{D}_\nu \fstarup =0,
\dq
where $\tilde{D}_\nu$ is the covariant derivative corresponding not
to $A_\nu$ but to a dual potential.
Secondly, the Yang--Mills equation, unlike
its abelian counterpart (\ref{freemax2}), says nothing about whether
the 2-form ${}^*\!F$ is closed or not.  Nor is the relation
(\ref{ymfcurl}) about exactness at all.  In other words, the Yang--Mills
equation does not guarantee the existence of a dual potential, in
contrast to the Maxwell case.   In fact, Gu and Yang \cite{Guyang}
have constructed a counter-example.  Because the true variables of
a gauge theory are the potentials and not the fields, this means
that Yang--Mills theory is {\em not symmetric} under the Hodge star
operation (\ref{hodge}) which in the abelian case gives us the duality
transform.

Nevertheless, electric--magnetic duality is a very useful physical
concept.  So one may wish to seek a more general duality
transform $(\tilde{\ })$ satisfying the following properties:
\begin{enumerate}
\item $(\quad)^{\sim\sim} = \pm (\quad)$,
\item electric field $F_{\mu\nu} \stackrel{\sim}{\longleftrightarrow}
$ magnetic field $\tilde{F}_{\mu\nu}$,
\item both $A_\mu$ and $\tilde{A}_\mu$ exist as potentials (away from
charges),
\item magnetic charges are monopoles of $A_\mu$, and electric charges
are monopoles of $\tilde{A}_\mu$,
\item $\tilde{\ }$ reduces to ${}^*$ in the abelian case.
\end{enumerate}

One way to do so is to study the Wu--Yang criterion more closely.
This reveals the concept of charges as topological constraints to be
crucial even in the pure field case, as can be seen in the diagram below:
$$
\begin{array}{ccc}
\fbox{\shortstack{$A_\mu$ exists as \\ potential for $F_{\mu\nu}$ \\
$[\,F={\rm d}\,A\,]$}} &
\stackrel{{\rm Poincar\acute{e}}}{\Longleftrightarrow}
& \fbox{\shortstack{Defining constraint \\$\partial_\mu
\fstarup=0$ \\ $[\,{\rm d}\,F =0\,]$}} \\
&&\\
\Big\Updownarrow & & \Big\Updownarrow\vcenter{%
\rlap{$\scriptstyle{\rm Gauss}$}}\\
&& \\
\fbox{\shortstack{Principal $A_\mu$ \\ bundle trivial}}  &
 &
\fbox{\shortstack{No magnetic\\ monopole $\tilde{e}$}}\\
&&\\
{\rm GEOMETRY} && {\rm PHYSICS}
\end{array}
$$
The point to stress is that, in the above abelian case, the condition
for the absence of a topological charge (a monopole) exactly removes
the redundancy of the variables \fdown, and hence recovers the
potential $A_\mu$.

Now the nonabelian monopole charge was defined topologically as an
element of $\pi_1 (G)$, and this definition also holds in the abelian
case of $U(1)$, with $\pi_1(U(1)) = \bbz$.  So the first task is to
write down a condition for the absence of a nonabelian monopole.

Consider the gauge invariant Dirac phase factor (or holonomy)
$\Phi (C)$ of a loop $C$, which can be written symbolically
as a path-ordered exponential:
\bq
\Phi[\xi] = P_s \exp ig \int_0^{2\pi} ds\,A_\mu(\xi(s))
\,\dot{\xi}^\mu(s), \label{phixis}
\dq
where we parametrize the loop $C$:
\bq
C: \ \ \ \{\xi^\mu(s) \colon s = 0 \rightarrow 2\pi,\ \xi(0) = \xi(2\pi)
   = \xi_0 \},
\dq
and a dot denotes differentiation with respect to the parameter $s$.
We thus regard loop variables in general as functionals of continuous
piecewise smooth functions $\xi$ of $s$.  In this way, loop
derivatives and loop integrals are just functional derivatives and
functional integrals.   This means that loop derivatives $\delta_\mu
(s)$ are defined by a regularization procedure approximating delta
functions with finite bump functions and then taking limits in a
definite order.

\begin{figure}
\centering
\input{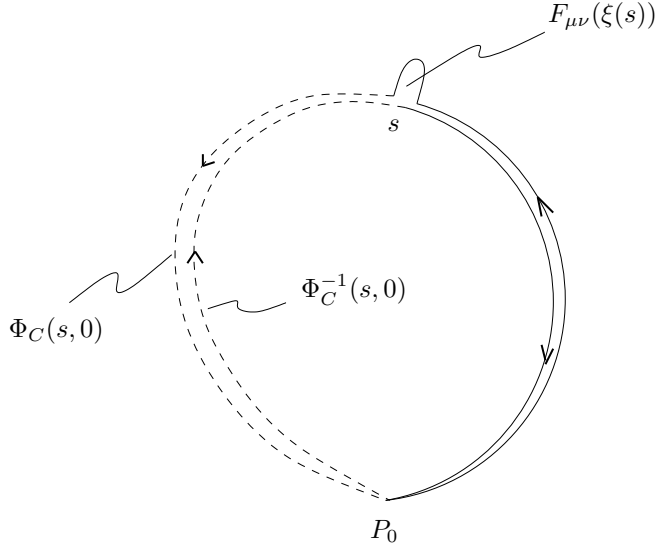}
\caption{Illustration for `loop connection'.}
\label{pluckloop}
\end{figure}

Following Polyakov~\cite{Polyakov} we introduce the logarithmic
loop derivative of
$\Phi[\xi]$:
\bq
F_\mu[\xi|s] = \frac{i}{g} \,\Phi^{-1}[\xi]\, \delta_\mu(s) \,\Phi[\xi],
\label{fmu}
\dq
which acts as a kind of `connection' in loop space since it tells us
how the phase of $\Phi[\xi]$ changes from one loop to a neighbouring
loop, as illustrated in Figure~\ref{pluckloop}.
One can go a step further and define its `curvature' in direct analogy
with $\fdown (x)$:
\bq
G_{\mu\nu}[\xi|s] = \delta_\nu(s) F_\mu[\xi|s]- \delta_\mu(s) F_\nu[\xi|s]
           + ig \,[\,F_\mu[\xi|s], F_\nu[\xi|s]\,].
\dq

It can be shown that using the $F_\mu[\xi|s]$ we can rewrite the
Yang--Mills action as
\bq
{\cal A}_F^0 = -\frac{1}{4 \pi \bar{N}} \int \delta \xi \int_0^{2\pi} ds
  \,{\rm Tr}\{F_\mu[\xi|s] F^\mu[\xi|s] \} \,|\dot{\xi}(s)|^{-2},
\label{loopaction}
\dq
where the normalization factor $\bar{N}$ is an infinite constant.
However, the true variables of the theory are still the $A_\mu$.  They
represent 4 functions of a real variable, whereas the loop connections
represent 4 functionals of the real function $\xi (s)$.  Just as in
the case of the \fdown, these $ F_\mu[\xi|s]$ have to be constrained
so as to recover $A_\mu$,
but this time much more severely.

It turns out that in pure Yang--Mills theory, the constraint that says
there are no monopoles:
\bq
G_{\mu\nu}[\xi|s] = 0 \label{loopcurv0}
\dq
removes also the redundancy of the loop variables, exactly as in the
abelian case.   That this condition is necessary is easy to see, by
simple algebra.  The proof of the converse of this ``extended
Poincar\'e lemma'' \cite{Chanstsou1,ourbook} is fairly lengthy and
will not be presented.  Granted this, we can now apply the Wu--Yang
criterion to the action (\ref{loopaction})
and derive the Polyakov equation:
\bq
\delta_\mu(s) F^\mu[\xi|s] = 0, \label{polyakov}
\dq
which is the loop version of the Yang--Mills equation.

In the presence of a  monopole charge $-$, if we use the $SO(3)$
example as an illustration,
the constraint
(\ref{loopcurv0}) will have a nonzero right hand side:
\bq
G_{\mu\nu}[\xi|s] = - J_{\mu\nu}[\xi|s].
\dq
The loop current $J_{\mu\nu}[\xi|s]$ can be written down explicitly.
However, its global form is much easier to understand.
Recall that $F^\mu[\xi|s]$ can be thought of as
a loop connection, for which we can form {\em its} `holonomy'.  This
is defined for a closed (spatial) surface $\Sigma$ (enclosing the
monopole), parametrized by a family of closed curves $\xi_t (s),\ t=0
\to 2\pi$.  The `holonomy' $\Theta_\Sigma$ is then the total change
in phase of
$\Phi[\xi_t]$ as $t \to 2 \pi$, and thus equals the charge $-$.

\begin{figure}
\centering
\input{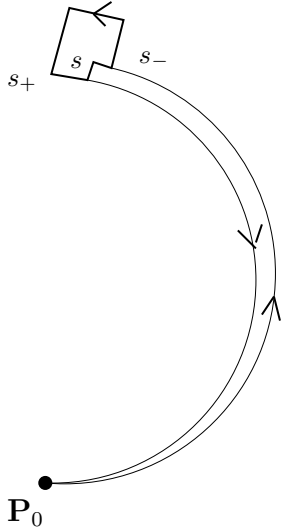}
\caption{Illustration of the segmental variable $E_\mu$.}
\label{segment}
\end{figure}

To formulate an electric--magnetic duality which is applicable to
nonabelian theory one defines yet another set of loop variables.
Instead of the Dirac phase factor $\Phi [\xi]$ for a complete curve
(\ref{phixis}) we consider the parallel phase transport for part of
a curve from $s_1$ to $s_2$:
\bq
\Phi_\xi(s_2,s_1) = P_s \exp ig \int_{s_1}^{s_2} ds\, A_\mu(\xi(s))
   \,\dot{\xi}^\mu (s).
\dq
Then the new variables are defined as:
\bq
E_\mu[\xi|s] = \Phi_\xi(s,0) \,F_\mu[\xi|s] \,\Phi_\xi^{-1}(s,0).
\dq
These are not gauge invariant like $F_\mu[\xi|s]$ and may not be as
useful in general but seem more convenient for dealing with duality.
A schematic representation is given in Figure~\ref{segment}.

Using these variables, we now define \cite{dualsymm} their
{\em dual} $\tilde{E}_\mu[\eta|t]$ as:
\begin{eqnarray}
& & \omega^{-1}(\eta(t)) \,\tilde{E}_\mu[\eta|t] \,\omega(\eta(t))
\nonumber\\
&=& -\frac{2}{\bar{N}} \,\epsilon_{\mu\nu\rho\sigma} \,\dot{\eta}^\nu(t)
     \int\!\delta \xi \,ds\, E^\rho[\xi|s] \,\dot{\xi}^\sigma(s)
\,\dot{\xi}^{-2}(s) \,\delta(\xi(s) -\eta(t))\,,
\label{eduality}
\end{eqnarray}
where $\omega(x)$ is a (local) rotation matrix transforming from the frame in
which the orientation in internal symmetry space of the fields $E_\mu[\xi|s]$
are measured to the frame in which the dual fields $\tilde{E}_\nu[\eta|t]$
are measured.  It can be shown that this dual transform satisfies all
the 5 required conditions we listed before.

Electric--magnetic duality in Yang--Mills theory is now fully
re-established using this generalized duality.  We have the following
dual pairs of equations:
\begin{eqnarray}
\delta_\nu E_\mu - \delta_\mu E_\nu &=& 0, \label{exista}\\
\delta^\mu E_\mu &=& 0; \label{eym}
\end{eqnarray}
and dually
\begin{eqnarray}
\delta^\mu \tilde{E}_\mu &=& 0, \label{eymdual}\\
\delta_\nu \tilde{E}_\mu - \delta_\mu \tilde{E}_\nu &=& 0. \label{existadual}
\end{eqnarray}

Equation (\ref{exista}) guarantees that the potential $A$ exists, and
so is equivalent to (\ref{loopcurv0}), and hence is  the nonabelian
analogue of (\ref{freemax1}); while equation (\ref{eym}) is equivalent
to the Polyakov version of Yang--Mills equation (\ref{polyakov}), and
hence is the nonabelian analogue of (\ref{freemax2}).  Equation
(\ref{eymdual}) is equivalent by duality to (\ref{exista}) and
is the dual Yang--Mills equation.
Similarly equation (\ref{existadual}) is equivalent to (\ref{eym}),
and guarantees the existence of the dual potential $\tilde{A}$.

The treatment of charges using the Wu--Yang criterion
also follows the abelian case, and will not be
further elaborated here.  For this and further details the reader is
referred to the original papers \cite{dualsymm}.

Also just as in the abelian case, the gauge symmetry is doubled: from
the group $G$ we deduce that the full gauge symmetry is in fact
$G \times \tilde{G}$, but that the physical degrees of freedom remain the
same.

\setcounter{equation}{0}

\section{The Dualized Standard Model}

That duality exists also for nonabelian Yang-Mills theory, as outlined in
the last section, is of basic theoretical interest, and is likely to have
repercussions in many areas of physics.  So far, however, the applications
we have made are concentrated in the problem of fermion generations where,
as we shall see, the consequences are both concrete and immediate.  The
reason is as follows.

We recall that according to present experiment, fermions occur in 3 and
apparently only 3 generations, which suggests a hidden 3-fold symmetry,
known in the literature for historical reasons as ``horizontal symmetry''
\cite{horizontal}.  This symmetry must be broken, and in a rather unusual
manner, given the peculiar hierarchical fermion mass spectrum quoted
above (\ref{masses}).  In most previous studies, the existence as well
as the breaking of this horizontal symmetry have to be taken as inputs
thus reducing the overall predictive power.  But with the nonabelian
duality derived above, the idea takes on a more concrete shape.  First,
dual to colour $SU(3)$, one knows that there is automatically another,
dual colour symmetry $\widetilde{SU}(3)$ bearing a similar relationship
to colour as magnetism bears to electricity.  Secondly, one knows that
this dual colour symmetry is broken.  This follows from a result of
't~Hooft \cite{tHooft} which says that if colour $SU(3)$ is confined, as
it is, then its dual is necessarily broken.  Indeed, using the machinery
developed in Section 2, it can be shown \cite{dualcomm} that the Wilson
operators:
\begin{equation}
A(C) = {\rm Tr} \left[P \exp ig \oint_C A_i(x) dx^i \right],
\label{AC}
\end{equation}
and:
\begin{equation}
B(C) = {\rm Tr} \left[ P \exp i \tilde{g} \oint_C \tilde{A}_i(x) dx^i \right],
\label{BC}
\end{equation}
constructed from respectively the colour potential $A_i(x)$ and the dual
colour potential $\tilde{A}_i(x)$, satisfy the commutation relation used
by 't~Hooft to derive his result, namely:
\begin{equation}
A(C) B(C') = B(C') A(C) \exp(2\pi i l/N)
\label{ABcomm}
\end{equation}
for $SU(N)$ gauge group and
for any 2 spatial loops $C$ and $C'$ with linking number $l$.

In other words, this means that, by virtue of nonabelian duality, there
is within the Standard Model framework already hidden a broken 3-fold
symmetry corresponding to dual colour which can play the role of the
horizontal symmetry for generations.  That being the case, it seems
natural to identify dual colour $\widetilde{SU}(3)$ as such.  Indeed,
if one does not do so, one may be at a loss as to what physical significance
to assign to this symmetry which, according to nonabelian duality, will
be there in any case.  But with this identification, one may claim that
nonabelian duality predicts the existence of 3 and only 3 generations
as experimentally observed.

The cited result of 't~Hooft shows that the $\widetilde{SU}(3)$ dual
colour symmetry is broken, but offers no hint as to the Higgs mechanism
for breaking it.  Interestingly, however, the framework developed above
for nonabelian duality itself suggests natural candidates for the Higgs
fields.  It was noted before \cite{dualsym} that the transformation
matrix $\omega(x)$ relating the colour to dual colour frame which appears
in the dual transform (\ref{eduality}) has to be patched (or
alternatively
to carry a Dirac string) in the presence of charges, and in monopole
theory, according to Wu and Yang \cite{WuYang}, it is the patching in
gauge fields in the presence of charges which gives rise to interactions
between them.  Hence, the observation that $\omega(x)$ be patched suggests
that its elements, or else the colour and dual colour frame vectors from
which it is constructed, can play a dynamical role and be considered for
promotion to physical fields.  The idea of promoting frame vectors to be
physical fields is of course not new, a well-known previous example being
the ``vierbeins'' in the Einstein-Cartan-Kibble-Sciama formulation of
relativity \cite{Heyl}.  If one were to promote the dual colour frame
vectors to fields, then they would have the appropriate properties of
the Higgs fields necessary for breaking the dual colour symmetry, being
triplets of dual colour, space-time scalars and having finite ``classical''
lengths.

The starting assumption of our Dualized Standard Model scheme is then to
make the identifications of dual colour to generations and of frame vectors
to Higgs fields for breaking the dual colour symmetry.  Apart from the
practical advantages to be detailed below, this has, to us, the aesthetic
appeal of assigning to both generations and Higgs fields a geometric
significance which they so sadly lack in our conventional formulation of
the Standard Model.

To proceed further, one needs an action, in particular the couplings of the
Higgs fields, i.e.\ the dual colour frame vectors, first to themselves and
second to the fermions.  The dual colour frame vectors represent 3 complex
triplet scalar fields $\phi^{(a)}_a, (a) = 1,2,3$, where $a = 1,2,3$ are
dual colour or generation indices.  Being frame vectors, and therefore
having equal status, they ought, we argued \cite{physcons}, to appear in
the action symmetrically.  We sought thus to construct with these a Higgs
potential which is renormalizable, $\widetilde{SU}(3)$ invariant, and
symmetric under permutations of the 3 triplets but having a degenerate
vacuum which breaks both the dual colour $\widetilde{SU}(3)$ symmetry and
the permutation symmetry spontaneously.  We proposed in \cite{physcons}
the following:
\begin{equation}
V[\phi] = -\mu \sum_{(a)} |\phi^{(a)}|^2 + \lambda \left\{ \sum_{(a)}
   |\phi^{(a)}|^2 \right\}^2 + \kappa \sum_{(a) \neq (b)} |\bar{\phi}^{(a)}
   .\phi^{(b)}|^.
\label{Vofphi}
\end{equation}
It has degenerate vacua of the form:
\begin{equation}
\phi^{(1)} = \zeta \left( \begin{array}{c} x \\ 0 \\ 0 \end{array}
\right), \ \ 
\phi^{(2)} = \zeta \left( \begin{array}{c} 0 \\ y \\ 0 \end{array}
\right), \ \ 
\phi^{(3)} = \zeta \left( \begin{array}{c} 0 \\ 0 \\ z \end{array} \right),
\label{phivac}
\end{equation}
with
\begin{equation}
\zeta = \sqrt{\mu/2\lambda},
\label{zeta}
\end{equation}
and $x, y, z$ all real and positive, satisfying:
\begin{equation}
x^2 + y^2 + z^2 = 1,
\label{xyznorm}
\end{equation}
which breaks the permutation symmetry between the $\phi$'s, and also the
$\widetilde{SU}(3)$ gauge symmetry completely.  In fact, all 9 (dual)
gauge bosons in the theory acquire a mass, eating up all but 9 of the
original 18 real Higgs modes.

Further, by analogy to the electroweak theory we proposed \cite{physcons}
the following Yukawa coupling to the fermions fields, again symmetric
under permutations of the 3 Higgs triplets:
\begin{equation}
\sum_{(a)[b]} Y_{[b]} \bar{\psi}_L^a  \phi^{(a)}_a \psi_R^{[b]} +
{\rm h.c.},
\label{Yukawa}
\end{equation}
where $\psi_L^a,\ a = 1,2,3$ is the left-handed fermion field appearing as a
dual colour triplet, and $\psi_R^{[b]}$, $[b]=1,2,3$,
are 3 right-handed fermion fields,
each appearing as a dual colour singlet\footnote{We note that in order to
have dual colour triplets occurring as monopoles of colour as we do here,
the colour $SU(3)$ group has to be imbedded in a larger theory as indeed
it is in the Standard Model.  This also makes it possible to have 9
gauge bosons acquiring mass, as stated above.
For a detailed explanation of this point, see
e.g.\ \cite{physcons}.}.

Neither the Higgs potential (\ref{Vofphi}) above nor the Yukawa coupling
(\ref{Yukawa}) can claim to be unique as implementations of the duality
ideas introduced before, and have thus to be regarded at present as
phenomenological constructs pending justification on a more theoretical
basis, which we have some hope of supplying in future but have not yet
succeeded in doing so.  They may thus possibly be subject to modifications.  
However, although the successes we shall show later in reproducing the 
fermion mass and mixing patterns have been obtained with these explicit 
constructs, we shall see indications that the most salient features could 
probably be retained under more general conditions.

 For the moment, however, let us continue with the explicit constructs
(\ref{Vofphi}) and (\ref{Yukawa}) and explore the consequences.  First, by
inserting the vacuum expectation values (\ref{phivac}) of the Higgs fields
$\phi^{(a)}_a$ into the Yukawa coupling (\ref{Yukawa}), one obtains the
fermion mass matrix at tree level:
\begin{equation}
\tilde{m} \frac{1}{2} (1 + \gamma_5) + \tilde{m}^{\dagger} \frac{1}{2}
   (1 - \gamma_5),
\label{mtreep}
\end{equation}
where $\tilde{m}$ is a factorized matrix:
\begin{equation}
\tilde{m} = \zeta \left( \begin{array}{c} x \\ y \\ z \end{array} \right)
   (a, b, c),
\label{mtree}
\end{equation}
with $a, b, c$ being the Yukawa couplings $Y_{[b]}$.  For future discussion
it is convenient, following Weinberg \cite{Weinberg}, to rewrite the mass
matrix $\tilde{m}$ in a hermitian form, basically replacing $\tilde{m}$ by
$\sqrt{\tilde{m} \tilde{m}^{\dagger}}$.  This can always be done by a
relabelling of the right-handed singlet fields $\psi_R^{[b]}$ without in
any way affecting the physics, as will be explicitly demonstrated for a
general mass matrix in the next section.  Applied to $\tilde{m}$ above,
one obtains:
\begin{equation}
m = m_T \left( \begin{array}{c} x \\ y \\ z \end{array} \right) (x, y, z),
\label{mtreew}
\end{equation}
which gives the physical states directly as the mass eigenstates.

We note first that apart from the proportionality factor $m_T$ for
$T = U, D, L, N$, this tree-level mass matrix is the same for all the 4
fermion species, which means in particular that at tree-level the up and
down mass matrices are aligned, hence giving no mixing at zeroth order,
which is no bad approximation at least for quarks.  Secondly, we note that
this matrix is of rank 1, having thus only one nonzero eigenvalue, which
we may interpret as an embryonic version of fermion mass hierarchy and
is a consequence of the stipulated condition that our action be invariant
under permutations of the three Higgs fields.  In other words, one begins
to see already the empirical fermion mass and mixing patterns taking shape.

One can do better, however.  Given the Higgs potential (\ref{Vofphi}) and
the Yukawa coupling (\ref{Yukawa}), it is only a matter of working through
some algebra \cite{ckm} to arrive at the following mass spectrum for the
remaining 9 Higgs bosons:
\begin{eqnarray}
K = 1: & 8 \lambda \zeta^2 (x^2+y^2+z^2), \nonumber \\
K = 2: & 4 \kappa \zeta^2 (y^2 + z^2)   , \nonumber \\
K = 3: & 4 \kappa \zeta^2 (y^2 + z^2)   , \nonumber \\
K = 4: & 4 \kappa \zeta^2 (z^2 + x^2)   , \nonumber \\
K = 5: & 4 \kappa \zeta^2 (z^2 + x^2)   , \nonumber \\
K = 6: & 4 \kappa \zeta^2 (x^2 + y^2)   , \nonumber \\
K = 7: & 4 \kappa \zeta^2 (x^2 + y^2)   , \nonumber \\
K = 8: & 0                              , \nonumber \\
K = 9: & 0                              ,
\label{MK1to9}
\end{eqnarray}
and the following for their couplings to fermions:
\begin{equation}
\bar{\Gamma}_K = \bar{\gamma}_K \frac{1}{2} (1 + \gamma_5)
   + \bar{\gamma}^{\dagger}_K \frac{1}{2} (1 - \gamma_5),
\label{Gammabar}
\end{equation}
where:
\begin{equation}
\bar{\gamma}_K = \rho |v_K \rangle \langle v_1|,
\label{gammabar}
\end{equation}
and:
\begin{eqnarray}
|v_1 \rangle & = & \left( \begin{array}{c} x \\ y \\ z \end{array} \right),
   \nonumber \\
|v_2 \rangle & = & \frac{1}{\sqrt{y^2 + z^2}}
   \left( \begin{array}{c} 0 \\ y \\ z \end{array} \right), \nonumber \\
|v_3 \rangle & = & \frac{i}{\sqrt{y^2 + z^2}}
   \left( \begin{array}{c} 0 \\ y \\-z \end{array} \right), \nonumber \\
|v_4 \rangle & = & \frac{1}{\sqrt{z^2 + x^2}}
   \left( \begin{array}{c} x \\ 0 \\ z \end{array} \right), \nonumber \\
|v_5 \rangle & = & \frac{i}{\sqrt{z^2 + x^2}}
   \left( \begin{array}{c} -x\\ 0 \\ z \end{array} \right), \nonumber \\
|v_6 \rangle & = & \frac{1}{\sqrt{x^2 + y^2}}
   \left( \begin{array}{c} x \\ y \\ 0 \end{array} \right), \nonumber \\
|v_7 \rangle & = & \frac{i}{\sqrt{x^2 + y^2}}
   \left( \begin{array}{c} x \\-y \\ 0 \end{array} \right), \nonumber \\
|v_8 \rangle & = & -\beta \left( \begin{array}{c} y-z \\ z-x \\
   x-y \end{array} \right), \nonumber \\
|v_9 \rangle & = & \beta \left( \begin{array}{c} 1-x(x+y+z) \\ 1-y(x+y+z) \\
   1-z(x+y+z) \end{array} \right),
\label{vK1to9}
\end{eqnarray}
with
\begin{equation}
\beta^{-2} = 3 - (x+y+z)^2.
\label{beta}
\end{equation}

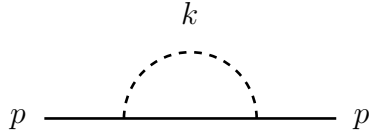
\begin{figure}[ht]
\begin{center}
{\unitlength=1.0 pt \SetScale{1.0} \SetWidth{1.0}
\begin{picture}(130,80)(0,0)
\Line(10,5)(120,5)
\DashCArc(65,5)(25,0,180){3}
\Text(0,5)[]{$p$}
\Text(130,5)[]{$p$}
\Text(65,45)[]{$k$}
\end{picture}}
\end{center}
\caption{1-loop insertion to the fermion propagator.}
\label{loopdiag}
\end{figure}

Given the above information, it is then possible to calculate the loop
corrections with these dual colour Higgs bosons exchanged, in particular
the 1-loop insertion of Figure \ref{loopdiag} to the fermion propagator
where the dashed line represents one of the Higgs boson states listed in
(\ref{MK1to9}).  Even at the 1-loop level, of course, there will be many
more diagrams giving insertions to the fermion propagator but these will
all be seen to yield but negligible contributions to calculating the
fermion mass and mixing patterns which is our main concern here and can
thus for the present be ignored.  For Figure \ref{loopdiag} then, one
has explicitly:
\begin{equation}
\Sigma(p) = \frac{i}{(4 \pi)^4} \sum_K \int d^4 k \;
   \frac{1}{k^2 - M_K^2}\;
   \bar{\Gamma}_K \,\frac{(p\llap/ - k\llap/) + m}{(p - k)^2 - m^2}
   \;\bar{\Gamma}_K,
\label{Sigma}
\end{equation}
with $m$ and $\bar{\Gamma}_K$ given in (\ref{mtreew}) and (\ref{Gammabar}).
Combining denominators by the standard Feynman parametrization and shifting
the origin of the $k$-integration as usual, one obtains:
\begin{equation}
\Sigma(p) = \frac{i}{(4 \pi)^4} \sum_K \int_0^1 dx \;\bar{\Gamma}_K
   \left\{ \int d^4 k \frac{p\llap/ (1 -x) + m}{[k^2 - Q^2]^2} \right\}
   \bar{\Gamma}_K,
\label{Sigma1}
\end{equation}
with
\begin{equation}
Q^2 = m^2 x + M_K^2 (1 - x) - p^2 x(1 -x),
\label{Qsquare}
\end{equation}
where we note that $m$, being a matrix in generation space, cannot be
commuted through the couplings $\bar{\Gamma}_K$.  The integration over
$k$ in (\ref{Sigma1}) is divergent and has to be regularized.   Following
the standard dimensional regularization procedure, one obtains:
\begin{equation}
\Sigma(p) = - \frac{1}{16 \pi^2} \sum_K \int_0^1 dx\; \bar{\Gamma}_K
   \{\bar{C} - \ln (Q^2/\mu^2) \} \{p\llap/ (1 - x) + m\}\; \bar{\Gamma}_K,
\label{Sigma2}
\end{equation}
with $\bar{C}$ being the divergent constant:
\begin{equation}
\bar{C} = \lim_{d \rightarrow 4} \left[ \frac{1}{2 - d/2} - \gamma \right],
\label{Cbar}
\end{equation}
to be subtracted in the standard $\overline{\rm MS}$ scheme.

To extract the renormalized mass matrix:
\begin{equation}
m' = m + \delta m
\label{mprime}
\end{equation}
from $\Sigma(p)$, one normally puts in the denominator $p^2 = m^2$ and
commutes $p\llap/$ in the numerator to the left or right and replace by
$m$ \cite{Weinberg}.  However, $m$ being now a matrix, this operation is
a little more delicate.  In order to maintain the ``hermitian'', left--right
symmetric form (\ref{mtreew}) for the renormalized mass matrix $m'$, we
split the $p\llap/$ term into two halves, commuting half to the left and
half to the right before replacing by $m$, and hence obtain for $\delta m$
the following:
\begin{eqnarray}
\delta m & = & \frac{\rho^2}{16 \pi^2} \sum_K \int_0^1 dx\;
   \{ \bar{\gamma}_K m\, [\bar{C} - \ln (Q_0^2/\mu^2)]\, \bar{\gamma}_K
      \,\frac{1}{2}(1 + \gamma_5) \nonumber \\
   & & {} + \bar{\gamma}_K^{\dagger} m\, [\bar{C} - \ln (Q_0^2/\mu^2)]\,
      \bar{\gamma}_K^{\dagger}\, \frac{1}{2}(1 - \gamma_5) \} \nonumber \\
   & & {} + \frac{\rho^2}{32 \pi^2} \sum_K \int_0^1 dx\, (1 - x)\,
      m\; \{ \bar{\gamma}_K^{\dagger}\,[\bar{C} - \ln (Q_0^2/\mu^2)]\,
      \bar{\gamma}_K\, \frac{1}{2}(1 + \gamma_5) \nonumber \\
   & & {} + \bar{\gamma}_K\,[\bar{C} - \ln (Q_0^2/\mu^2)]\,
      \bar{\gamma}_K^{\dagger}\,\frac{1}{2}(1 - \gamma_5) \} \nonumber \\
   & & {} + \frac{\rho^2}{32 \pi^2} \sum_K \int_0^1 dx\,(1 - x)\;
      \{ \bar{\gamma}_K\,[\bar{C} - \ln (Q_0^2/\mu^2)]\,
      \bar{\gamma}_K^{\dagger}\,\frac{1}{2}(1 + \gamma_5) \nonumber \\
   & & {} + \bar{\gamma}_K^{\dagger}\,[\bar{C} - \ln (Q_0^2/\mu^2)]\,
      \bar{\gamma}_K\, \frac{1}{2}(1 - \gamma_5) \}\; m,
\label{deltam}
\end{eqnarray}
with
\begin{equation}
Q_0^2 = Q^2|_{p^2 = m^2} = m^2 x^2 + M_K^2 (1 - x).
\label{Qzerosquare}
\end{equation}
Again a relabelling of the right-handed fermion fields is required to bring
the renormalized mass matrix back to the hermitian form (\ref{mtreew}) of
Weinberg adopted here.

The expression (\ref{deltam}) for the 1-loop correction to the mass matrix
is a little complicated, but for the consideration of the fermion mass and
mixing patterns of main concern in this paper, the only relevant terms in
(\ref{deltam}) are those proportional to $\ln \mu^2$, with $\mu$ being
the renormalization scale.  The reason is that the remainder can readily
be shown \cite{ckm,transmudsm} to be of order $m^2/M^2$, where $M$ is a
mass scale bounded by present experimental limits on flavour-violation to
be of order 100 TeV \cite{ckm,fcnc}, to which questions we shall return at
the end of these lectures in section 7.  Keeping then only these $\ln \mu^2$
terms and summing over all the Higgs bosons labelled by $K$, one obtains
\cite{ckm}:
\begin{equation}
m' = m_T' \left( \begin{array}{c} x' \\ y' \\ z' \end{array} \right)
   (x', y', z'),
\label{mtreewp}
\end{equation}
where the vector $(x', y', z')$ satisfies an RG-type equation of the
following form:
\begin{equation}
\frac{d}{d(\ln \mu^2)} \left( \begin{array}{c} x' \\ y' \\ z'
   \end{array} \right)
   =  \frac{3}{64 \pi^2} \rho^2 \left( \begin{array}{c}
         x'_1 \\ y'_1 \\ z'_1 \end{array} \right),
\label{runxyz}
\end{equation}
with
\begin{equation}
x'_1 = \frac{x'(x'^2-y'^2)}{x'^2+y'^'2} + \frac{x'(x'^2-z'^2)}
   {x'^2+z'^2}, \ \ \ {\rm cyclic},
\label{x1tilde}
\end{equation}
and $\rho$ being the Yukawa coupling strength\footnote{There was an error
in \cite{ckm} which gave the coefficient on the right of eq.\ (\ref{runxyz})
as $5/(64 \pi^2)$ instead of $3/(64 \pi^2)$ as in here.
This was due to a sign
error in the first term on the right of eq.\ (4.14) of \cite{ckm} arising
from a misprint in the formula for $\Sigma^{(\phi 1)}$ in eq.\ (3.2)
of \cite{Weinberg} quoted there.  However, apart from the fact
that the numerical values
given for the parameter $\rho$ in eq.\ (6.8) in \cite{ckm}
should be increased
by a factor $\sqrt{5/3}$, no other results given in that paper or in its
sequels such as \cite{phenodsm} are affected by this error.}.

We notice first that the renormalized mass matrix (\ref{mtreewp}) remains
of the factorized form.  This result is independent of whether terms
of order $m^2/M^2$ are included or not
and will hold even with the inclusion of diagrams other
than the calculated Higgs loop of Figure \ref{loopdiag}.  It holds simply
by virtue of the also factorized form of the Higgs coupling as deduced from
(\ref{Yukawa}), and of the fact that the dual colour gauge bosons couple
only to left-handed fermions which are dual colour triplets but not to
right-handed fermions which are dual colour singlets \cite{ckm}.  Secondly,
we note that to the very good approximation of neglecting quantities of
order $m^2/M^2$, the vectors $(x', y', z')$ are identical for all 4 fermion
species $U, D, L, N$, so that the mass matrices are still the same apart
from the normalisation $m_T'$. (In principle, the Yukawa coupling strength
$\rho$ appearing in equation (\ref{runxyz}) can also depend on the fermion
species, but for reasons of consistency \cite{ckm,bordsm} to be reviewed
later, they have to be equal in the DSM scheme.)  The resulting picture is
thus extremely simple, and formally similar to that at tree level.  There
is, however, a very important difference, namely that, in contrast to the
tree-level mass matrix (\ref{mtreew}), the vector $(x', y', z')$ factored
from the renormalized mass matrix is no longer constant but depends on
scale via the equations (\ref{runxyz}) and (\ref{x1tilde}).  It changes
not only in length but also in direction, which means that the mass matrix,
apart from running in normalization, also changes in orientation,
that is, rotates,
with changing scale.  And this difference, as we shall see in the next
section, is enough not only to give nontrivial mixing and nonzero masses to
the lower generations, both of which were missing in the tree approximation,
but also to offer an immediate explanation for almost all the salient
features of the experimentally observed fermion mass and mixing patterns
quoted in section 1, which had seemed so mysterious before.

\setcounter{equation}{0}

\section{The Rotating Mass Matrix and its Implications}

That the renormalized mass matrix should change with scale, like the coupling
constant and other field quantities, is of course no surprise, and that it
should rotate also is not peculiar just to the DSM scheme but happens already
in the Standard Model as conventionally formulated \cite{Ramon,physcons},
although the rotation there is very weak and its effects are thus for most
applications negligible.  What is perhaps not widely recognized, however,
is that when the mass matrix does rotate, then some of our usual kinematical
concepts such as particle masses, state vectors and mixing parameters will
have to be refined.  This is a matter of principle which will have to be
faced in whatever situation where the mass matrix rotates, however weakly,
not just in the DSM scheme being considered.

The situation being unfamiliar, it would be worthwhile to examine it afresh
starting from basics and in terms of a general rotating mass matrix before
specializing later to the DSM case.  Let us start then with a fermion mass
matrix traditionally defined by a term in the action of the form:
\begin{equation}
\bar{\psi}^0_L\, {\tilde m}\, \psi^0_R\, + \,{\rm h.c.},
\label{mtrad}
\end{equation}
where $\psi^0_L$ and $\psi^0_R$ represent respectively the left- and
right-handed fermion field, each being a vector in 3-dimensional flavour
space, here given in the weak gauge basis, and ${\tilde m}$ is a $3 \times 3$
(complex) matrix.  The matrix ${\tilde m}$ can always be diagonalized as
follows:
\begin{equation}
U_L^\dagger\, {\tilde m}\, U_R = {\rm diag}\,\{m_1, m_2, m_3\}
\label{diagmtrad}
\end{equation}
with $U_L, U_R$ unitary and $m_i$ taken real.  Thus in terms of the fields:
\begin{equation}
\psi_L = U_L^\dagger\, \psi^0_L; \ \ \ \psi_R = U_R^\dagger\, \psi^0_R,
\label{mtradev}
\end{equation}
the term (\ref{mtrad}) in the action takes on the diagonal form:
\begin{equation}
\bar{\psi}_L \;{\rm diag}\,\{m_1, m_2, m_3\}\; \psi_R.
\label{mtraddiag}
\end{equation}
When the mass matrix ${\tilde m}$ is constant in orientation with respect
to scale change, i.e.\ in our language here, when the mass matrix does not
rotate, which is the simple case usually considered, then the particle
masses of the 3 flavour states are just given by the diagonal values $m_i$.
The above apply to both up and down quarks in the case of quarks, and to
both charged leptons and neutrinos in the case of leptons.  Hence, from
the mass matrix, one obtains for the up and down states each a diagonalizing
matrix $U_L$ which we can denote respectively as $U_L$ and $U'_L$.  Again,
in the simple case when the mass matrices do not rotate, then the mixing
matrix between up and down states (i.e.\ CKM \cite{CKM} for quarks and MNS
\cite{MNS} for leptons) is just given by \cite{Jarlskog}:
\begin{equation}
V = U_L\, {U'}^\dagger_L.
\label{mixingmJ}
\end{equation}

 For our discussion here, as mentioned already in (\ref{mtreew}), it is
more convenient to work with an equivalent form of the mass matrix adopted
by Weinberg in \cite{Weinberg}.  Since the right-handed fermion fields are
flavour singlets, they can be arbitrarily relabelled without changing any
of the physics. This is clear from
the fact that the mixing matrices between up and
down states depend only on $U_L$ and
not on $U_R$.  Hence, by an appropriate
relabelling of right-handed fields, explicitly by defining new right-handed
fields:
\begin{equation}
{\psi'}^0_R = U_L U_R^\dagger\, \psi^0_R,
\label{newpsir}
\end{equation}
one obtains (\ref{mtrad}) in a form in which the mass matrix becomes
hermitian:
\begin{equation}
\bar{\psi} m \half(1 + \gamma_5) \psi
   + \bar{\psi} m \half(1 - \gamma_5) \psi = \bar{\psi} m \psi,
\label{mtradW}
\end{equation}
with
\begin{equation}
m = {\tilde m}\, U_R U_L^\dagger.
\label{mW}
\end{equation}
This is convenient because in the simple case when the mass matrix does
not rotate, the particle masses are now just the real eigenvalues of the
hermitian matrix $m$ and the state vectors of flavour states just the
corresponding eigenvectors, as can readily be checked with (\ref{diagmtrad}).
Furthermore, the mixing matrix between up and down states becomes just
\begin{equation}
V_{ij} = \langle {\bf v}_i|{\bf v}'_j \rangle,
\label{ckmdot}
\end{equation}
with $|{\bf v}_i \rangle$
being the eigenvector of $m$ for the eigenvalue $m_i$
of the up state, and a prime denoting the corresponding quantities of the
down state.  In (\ref{ckmdot}), the scalar product
$\langle {\bf v}_i|{\bf v}'_j \rangle$
is of course an invariant independent of the frame in which these vectors
$|{\bf v}_i \rangle$ are expressed.

Consider now what happens in the case when the mass matrix
does rotate with
changing scale as is of interest to us here.  Both its eigenvalues and
their corresponding eigenvectors now change with the scale so that the
previous definition of these as respectively the masses and state vectors
of flavour states is no longer sufficiently precise, for it will have to
be specified at which scale(s) the eigenvalues and eigenvectors are to be
evaluated.

In the simple case of a single generation, i.e.\ when the mass matrix
is just a number, one is used to defining the particle mass as the running
mass taken at the scale equal to the mass value itself, i.e.\ at that $\mu$
at which $\mu = m(\mu)$.  Even in the multi-generation case when the mass
matrix does not rotate but its eigenvalues run with changing scales, one
can still define the mass $m_i$ and the state vector ${\bf v}_i$ of the
state $i$, as respectively just the $i$th eigenvalue and eigenvector of the
matrix $m$ taken at the scale $\mu_i = m_i(\mu_i)$, with $m_i(\mu)$ being
the scale-dependent $i$th eigenvalue of the matrix $m$.  One might therefore
be tempted to suggest the same definitions in the multi-generation case even
when the mass matrix rotates.  However, this will not do, because it would
mean that the state vectors for the different generations $i$ will be defined
as eigenvectors of the matrix $m$ at different scales.  Although
the eigenvectors $i$ for different eigenvalues $i$ are orthogonal, $m$
being hermitian,
when taken all at the same scale, they need not be mutually orthogonal when
taken each at a different scale.  But the state vectors for different flavour
states ought to be orthogonal to one another if they are to be independent
quantum states.  Otherwise, it would mean physically that the flavour states
would have nonzero components in each other and be thus freely convertible
into one another, or that the mixing matrices
would no longer be unitary, which would of course be unphysical.

How then should the mass values and state vectors of flavour states be
defined in the scenario when the mass matrix rotates?  To see how this
question may be resolved, let us examine it anew with first the $U$ type
quarks as example.  The $3 \times 3$ mass matrix $m$ has 3 eigenvalues
with the highest value $m_1$ corresponding to the eigenvector ${\bf v}_1$,
both depending on scale $\mu$.  Starting from a high scale and running
down, one reaches at some stage $\mu_1 = m_1(\mu_1)$, i.e.\ when the scale
equals the highest eigenvalue $m_1$.  One can then naturally define this
value $m_1(\mu_1)$ as the $t$ quark mass $m_t$ and the corresponding
eigenvector ${\bf v}_1(\mu_1)$ as the $t$ state vector ${\bf v}_t$.  Next,
how should one define the mass $m_c$ and the state vector ${\bf v}_c$?  We
have already seen above that they cannot be defined as respectively the
second highest eigenvalue $m_2$ of the $3 \times 3$ mass matrix $m$ and
its corresponding eigenvector at the scale $\mu_2 = m_2(\mu_2)$, because
this vector is in general not orthogonal to the state vector ${\bf v}_t$
which the state vector ${\bf v}_c$ ought to be.  It is not difficult,
however, to see what is amiss.  At scales below the $t$ mass, i.e.\ when
$\mu < m_t$, $t$ would no longer exist as a physical state, so that what
functions there as the fermion mass matrix is not the $3 \times 3$ matrix $m$
but only the $2 \times 2$ submatrix, say $\hat{m}$, of $m$ in the subspace
orthogonal to ${\bf v}_t$.  Hence, for consistency, one should define
$m_c$ as the highest eigenvalue $\hat{m}_2$ of the submatrix $\hat{m}$ and
the state vector ${\bf v}_c$ as the corresponding eigenvector, both at
the scale $\hat{\mu}_2 = \hat{m}_2(\hat{\mu}_2)$.  The state 
vector of $c$ so obtained is automatically orthogonal to ${\bf v}_t$ as
it should be.  Repeating the argument, one defines further the mass $m_u$
and state vector ${\bf v}_u$ respectively as the ``eigenvalue'' and
``eigenvector'' of $\hat{\hat{m}}$ at the scale $\hat{\hat{\mu}}_3 =
\hat{\hat{m}}_3(\hat{\hat{\mu}}_3)$, with $\hat{\hat{m}}$ being the
$1 \times 1$ submatrix of $m$ in the subspace orthogonal to both
${\bf v}_t$ and ${\bf v}_c$.  Proceeding in this way, all masses and state
vectors are defined at their own proper mass scale and the state vectors
are mutually orthogonal as they should be.  Besides, though stated above
only for 3, the definition can be extended to any number of fermion
generations, should there be physical incentive for doing so.

\begin{figure} [ht]
\vspace*{7mm}
\centering
\input{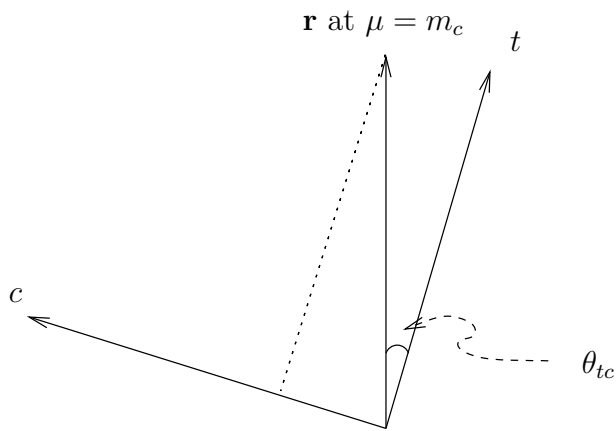}
\caption{Masses for lower generation fermions from a rotating mass matrix
via the ``leakage'' mechanism.}
\label{massleak}
\end{figure}

Having now made clear the general procedure for defining masses and state
vectors for a rotating mass matrix, let us return to consider in particular
the implications in the DSM scenario.  There, we recall in (\ref{mtreewp})
that the mass matrix is of a factorized form:
\begin{equation}
m = m_T |{\bf r} \rangle \langle {\bf r}|,
\label{mfact}
\end{equation}
given in terms of a single vector ${\bf r} = (x', y', z')$ which rotates
with changing scales and in which the whole content of the rotating mass
matrix is encapsulated.  Since our discussion depends only on the
orientation of this vector, the length of which cannot in any case at
present be calculated perturbatively, we shall henceforth take {\bf r}
to be a normalized vector.  This mass matrix
$m$ is of rank 1 and is aligned to a good
approximation for all fermion species.  Nevertheless, we claim that because
${\bf r}$ rotates with changing scale, we would obtain nonzero masses for
the lower generations as well as nontrivial mixing as a result.  This is
most easily seen by first considering the 2 heavier generations.
The procedure of the preceding paragraph 
gives  the state vector ${\bf v}_t$ of $t$ as
the single massive eigenstate ${\bf r}$ of the $U$ quark mass matrix at
the scale $\mu = m_t$.  As the scale lowers to $\mu=m_c$, the vector
${\bf r}$ will have rotated to a different direction as depicted in
Figure \ref{massleak}.  The state vector ${\bf v}_c$ is thus by
definition the vector orthogonal to ${\bf v}_t$ lying on the plane
spanned by ${\bf v}_t$ and ${\bf r} (m_c)$.
The $c$ mass $m_c$ is then given as the eigenvalue of
$\hat{m}$ at scale $\mu = m_c$, which for the rank 1 matrix $m$ in
(\ref{mfact}) 
is just the expectation value of $m$ in the state ${\bf v}_c$.  
Hence $c$  acquires by
``leakage'' a nonzero mass:
\begin{equation}
m_c = \langle {\bf v}_c|m|{\bf v}_c \rangle = m_t\, |\langle {\bf v}_c|
   {\bf r} \rangle|^2 = m_t\, \sin^2 \theta_{tc},
\label{planeleak}
\end{equation}
with $\theta_{tc}$ the rotation angle between the scales $\mu = m_t$
and $\mu = m_c$.

\begin{figure} [ht]
\centering
\input{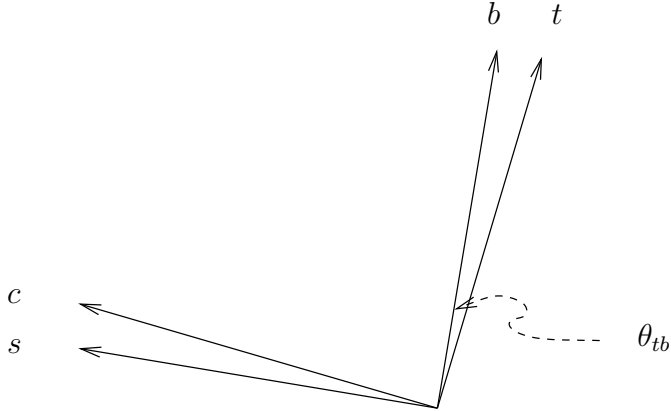}
\caption{Mixing between up and down fermions from a rotating mass matrix.}
\label{mixing}
\end{figure}

Similarly, although the mass matrices of the $U$ and $D$ quarks according
to (\ref{mtreewp}) are always aligned in orientation when both are at the
same scale, the state vectors ${\bf v}_t$ and ${\bf v}_b$ are defined as
the vector ${\bf r}$ at different scales, namely ${\bf v}_t = {\bf r}$
at $\mu = m_t$, but ${\bf v}_b = {\bf r}$ at $\mu = m_b$.  Hence, one
sees from Figure \ref{mixing} that simply by virtue of the rotation of
the vector ${\bf r}$ from the scale $\mu = m_t$ to the scale $\mu = m_b$,
a nonzero mixing between the $t$ and $b$ states results with the CKM matrix
element given by (\ref{ckmdot}) as:
\begin{equation}
V_{tb} = {\bf v}_t.{\bf v}_b = \cos \theta_{tb},
\label{planemix}
\end{equation}
where $\theta_{tb}$ is the rotation angle between the two scales.

Hence, already from these examples, one sees that both lower generation
masses and nontrivial mixing will automatically be obtained from the
rotating mass matrix (\ref{mtreew}) even if one starts with neither.  
Similar procedures apply to the lowest generation.

We conclude therefore that in spite of its simplicity the renormalized mass
matrix of (\ref{mtreewp}) is capable by virtue of the rotation induced by
equation
(\ref{runxyz}) of yielding nonzero masses for the lower generations
as well as nontrivial mixing between up and down fermion states.  The next
question then is whether it can give mass and mixing parameters to agree
in value with those observed in experiment.  Given the formalism already
set up above, it is in principle just a matter to be answered by performing
the suggested calculation, which has already been performed at the 1-loop
level and will be described in the next section.  However, before we do so,
it is worth examining the equations to familiarize ourselves with those
features which assure us of some reasonable answers.  Although we ourselves
learned to appreciate these only in hindsight after performing the said
calculations in detail, our job would have been much easier had we realised
them before.

To see this, let us examine the equation (\ref{runxyz}) in a little more
detail.  We note first from (\ref{x1tilde}) that the two points where
$(x', y', z')$ equals
$(1,0,0)$ or $\frac{1}{\sqrt{3}}(1,1,1)$ are rotational fixed points
for the vector.
Secondly, from (\ref{runxyz}) we see that as
the scale $\mu$ decreases, the vector ${\bf r} = (x', y', z')$ moves away 
from the point $(1,0,0)$ towards the point $\frac{1}{\sqrt{3}}(1,1,1)$.  
In other words, starting say at high scale, as the scale $\mu$ lowers, 
the vector ${\bf r}$ traces out a trajectory on the unit sphere joining 
the high energy fixed point $(1,0,0)$ to the low energy fixed point 
$\frac{1}{\sqrt{3}}(1,1,1)$.

\begin{figure}
\hspace*{-1.5cm}
\includegraphics[angle=-90, width=1.2\textwidth]{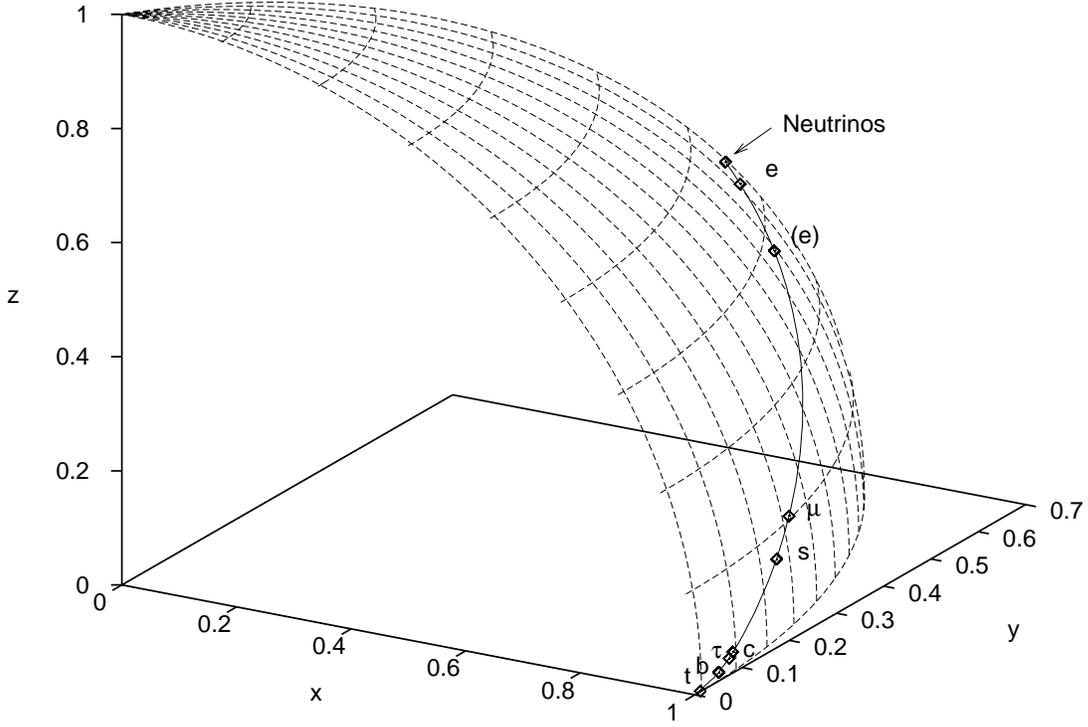}
\caption{Rotation trajectory of the vector ${\bf r} = (x', y', z')$ on 
the unit sphere as calculated in the 1-loop approximation of DSM in
\cite{phenodsm}.  The locations of the various fermions states marked 
on the trajectory represent their mass scales, thus for example, the 
location of $t$ is given by the scale $\mu = m_t = 175$ GeV.  For
the electron, we have marked 2 locations with $e$ corresponding to
$\mu = 0.51$ MeV, the empirical mass of the electron, and (e) to
$\mu = 6$ MeV, the calculated mass in the 1-loop approximation.
For all the other fermions marked except for the neutrinos, no such 
distinction is needed since the empirical mass and the calculated mass
coincide.  For neutrinos, the masses are so small and so close to the low 
energy fixed point $\frac{1}{\sqrt{3}}(1,1,1)$ as to be indistinguishable
in the figure.  From the marked locations of the various fermion states,
one can gauge the rotation speed of ${\bf r}$ with respect to change in 
scale $\mu$.  In particular, one notes that rotation is slow near either
of the 2 fixed points.}
\label{sphere}
\end{figure}

Near either fixed point, the rotation will of course be slower, and since
according to our previous discussion, both the leakage of masses to the
lower generations and the mixing between up and down states come in this
scheme from the rotation, so both these effects will also be smaller at
scales near the two fixed points.  Suppose therefore we were to choose
the parameters of the scheme so as to place the $t$ quark close
to the high energy fixed point $(1,0,0)$ but the neutrinos close to 
the low
energy fixed point $\frac{1}{\sqrt{3}}(1,1,1)$ as indicated 
in Figure \ref{sphere}.  (The trajectory in this figure is actually the
result of a calculation to be described later, but will serve 
as an illustration here.)  Then we would be able immediately to deduce the
following consequences.

(i) Since $t$ is nearer than $b$ to the fixed point $(1,0,0)$, and $b$
is nearer than $\tau$, the mass leakage will also go in that order, namely:
$m_c/m_t < m_s/m_b < m_\mu/m_\tau$, which agrees with the experimental
values quoted in (\ref{masses}).

(ii) Since the neutrinos are much further on the trajectory from the charged
leptons than the $D$ quarks are from the $U$ quarks, mixing angles are much
larger for leptons than for quarks.  This is again as seen in experiment as
quoted in (\ref{exckm}) and (\ref{exmns}).

(iii) With a bit more sophistication, using some elementary differential
geometry \cite{Docarmo}, it can be shown that the mixing matrices can be
approximately expressed in the form \cite{features}:
\begin{equation}
\left( \begin{array}{ccc}
       1 & -\kappa_g \Delta s & -\tau_g \Delta s \\
       \kappa_g \Delta s & 1 & \kappa_n \Delta s \\
       \tau_g \Delta s & -\kappa_n \Delta s & 1 \end{array} \right)
\label{ckmdg}
\end{equation}
to first order in the arc-length $\Delta s$ separating the heaviest up
state from the heaviest down state, where $\kappa_g$ is the geodesic
curvature, $\kappa_n$ the normal curvature, and $\tau_g$ the geodesic
torsion of the trajectory on a surface.  When the surface is the unit
sphere, as in our case, $\tau_g = 0$ and $\kappa_n = 1$.  This means
first that the corner elements of the mixing matrices, i.e.\ $V_{ub}$
and $V_{td}$ of CKM, and $U_{e3}$ of MNS, must be much smaller than the
other elements, which is seen to be the case in (\ref{exckm}) and
(\ref{exmns}).  Secondly, the 23 element is proportional roughly to
the separation $\Delta s$, which explains why the mixing angle 
$U_{\mu 3}$ for
atmospheric neutrinos is so much bigger than 
the corresponding angle $V_{cb}, V_{ts}$ for quarks, an experimental 
observation which has caused much recent excitement.

Thus, even without a detailed calculation, one can already see that there
is a good chance of obtaining qualitatively reasonable result from the
present scheme for fermion mass and mixing parameters.  The only question
is really whether one can choose the few parameters inherent in the scheme
to explain sufficiently the existing data.  This will
be decided by explicit calculations, which form the subject of the
next section.

Before we do so, however, we notice that in our above discussion from 
equation
(\ref{mfact}) onwards, we have taken the vector ${\bf r}$ factored from
the mass matrix to be a real vector to conform with what was obtained from
the DSM 1-loop calculation begun in the preceding section and continued in
the next.  This means that the CKM and MNS matrices which result are both
going to be real and can give no CP-violation.  But this is a limitation
only of the 1-loop calculation, not of the general considerations in this
section which can be repeated virtually unchanged with ${\bf r}$ complex,
thus accounting for a CP-violating phase, as might become necessary, for 
example, when higher loop effects are involved.

\setcounter{equation}{0}

\section{1-loop DSM Result on Masses and Mixing Angles}

The subject of this section is the calculation of the rotating fermion
mass matrix to 1-loop order with the initial object of explaining the
mass and mixing patterns of quarks and leptons as experimentally observed.
Since the parameters of the problem have yet to be determined by fitting
with data, the question of applicability and accuracy of the 1-loop
approximation, and if so in what physical range, can in principle only
be answered after the calculation has been performed, and then only to
the limit of our understanding.  However, anticipating our results, to
a discussion of which we shall return at the end, we suggest that the
calculation can be expected generally to be valid to a rough accuracy of
say 20 to 30 percent in mass ratios and mixing parameters over a range of
energy scales starting from about the top mass at 175 GeV down to about
the muon mass at 105 MeV.  As we shall see, however, there are special
circumstances which allow us to expect reasonable accuracy also for some
other quantities such as the elements $U_{e3}$ and $U_{\mu 3}$ of the
lepton mixing matrix associated with neutrino oscillations, although
these lie formally outside the above scale range.  These conclusions have
much to do with the existence of the two rotational fixed points mentioned
above at respectively infinite and zero scales, near to which the 1-loop
approximation has a better chance of being valid.

\begin{figure}[htb]
\vspace{1cm}
\centerline{\psfig{figure=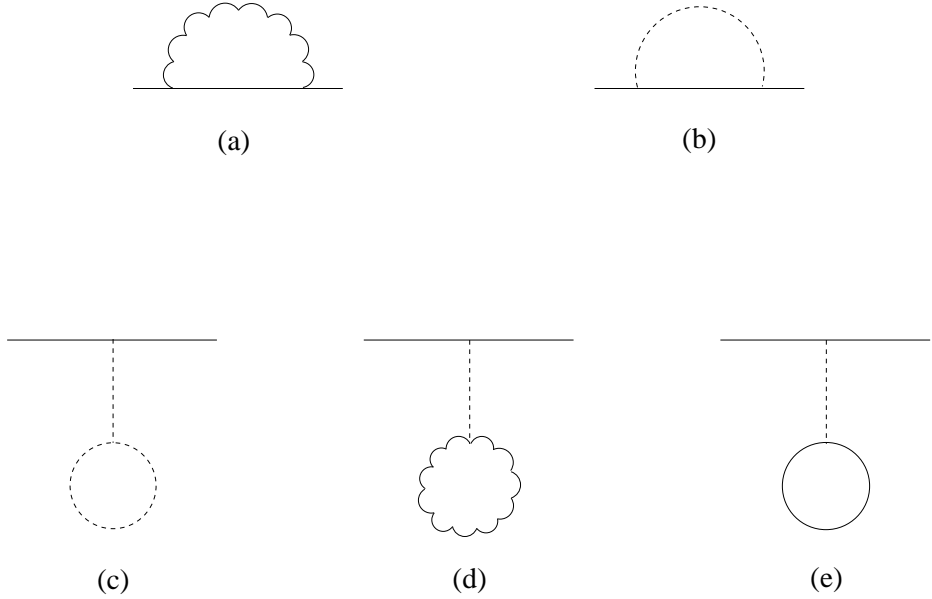,width=0.9\textwidth}}
\vspace{1cm}
\caption{One-loop diagrams with gauge and Higgs bosons carrying dual colour.}
\label{loopdiags}
\end{figure}

Even to 1-loop order, of course, there are in principle many diagrams which
can contribute to the renormalization of the fermion mass matrix.  However,
if we accept the contention made above, that mass ``leakages'' and mixings
of fermions are due mainly to mass matrix rotations, then the problem
simplifies tremendously \cite{ckm}.  First of all, the insertions of the
type present already in the conventional formulation of the Standard Model,
not being directly dependent on dual colour (i.e.\ the generation index)
contribute practically nothing to the rotation of the fermion mass matrix.
Secondly, of the new diagrams involving the exchange of gauge and Higgs
bosons carrying dual colour or generation index which are listed together
in Figure \ref{loopdiags}, all except the Higgs loop insertion already
calculated give rotations only of order $\mu^2/M^2$, with $M$ of order
100 TeV, and are therefore negligible for the effects we seek.
This is very fortunate, for it means that to 1-loop order, the result
already calculated and qualitatively analysed in the preceding 2 sections 
is all that we would need.

To the accuracy we need, then, the fermion mass matrix to 1-loop order is
given by (\ref{mtreewp}) in terms of a vector ${\bf r} = (x', y', z')$ in
3-dimensional generation space which rotates with changing scale according
to the evolution equation (\ref{runxyz}).  Referring back to (\ref{runxyz})
and (\ref{x1tilde}), one sees that apart from a mass scale $m_T$ for each
fermion species, the remaining freedom is only in the choice of trajectory
for the vector ${\bf r}$, and the Yukawa coupling strength $\rho$
which governs the speed of the vector's rotation along the trajectory.
The vector being by definition a unit vector, the trajectory will be
specified by a choice of some initial values, say $y_I, z_I$ of $y', z'$.
The coupling $\rho$ could, as mentioned already, depend on the
fermion species, but for consistency with the present interpretation have
to be the same for all.  This can be seen either numerically as shown in
\cite{ckm} or else analytically from an approximate solution of the
evolution equation \cite{bordsm}.  One has then altogether just 3 real
parameters to explain the mass ratios between generations and the mixing
matrices between up and down states for both quarks and leptons.

The equation (\ref{runxyz}), being linear, is easily integrated for
any given value of the coupling parameter $\rho$ and any initial point
on the trajectory, say $(x_I, y_I, z_I)$, at any chosen scale $\mu_I$.
The integration can be done numerically by iteration as in
\cite{ckm,phenodsm}, where for the 1 percent accuracy aimed for which would
be more than adequate for present purposes, roughly 500 steps of iteration
are made per decade change in energy, the vector ${\bf r} = (x', y', z')$
being re-normalized to unit length after every step.  Having obtained then
the trajectory, i.e.\ the vector ${\bf r}$ at every scale, it is an easy
matter, following the procedure described in the above section, and given
the normalization $m_T$ of the mass matrices, to calculate the mass ratios
between generations for each of the 4 fermion species $T = U, D, L, N$, as
well as the elements of both the CKM and MNS mixing matrix for quarks and
leptons respectively.  However, although for the up
and down quarks and charged leptons, the normalization $m_T$ can be taken
respectively as $m_t, m_b, m_\tau$ which are all by now quite well measured
\cite{databook}, $m_T$ for neutrinos is still unknown, being
given in the above formulation by the Dirac mass of the heaviest state $\nu_3$
(which is most likely distinct from the physical mass because of the see-saw
mechanism \cite{Ramon}).  Hence, only the numerical values of the mass
ratios for $T = U, D, L$ and the CKM mixing matrix for quarks are obtained
at this stage.  These numbers, however, still depend on the given choice of
parameters $y_I, z_I$ and $\rho$, and need not of course agree with
the empirical values.
One has thus first to determine the appropriate values of these
parameters by fitting to experiment.

 For convenience and without loss of generality, we can take
$x_I \ge y_I \ge z_I$ at some arbitrary high scale value $\mu_I$ which
is chosen to be 20 TeV in \cite{ckm,phenodsm}.  The strategy is then to fit
the 3 parameters $y_I, z_I$ and $\rho$ to the 3 best measured quantities
among the fermion mass and mixing parameters which are at present the mass
ratios $m_c/m_t, m_\mu/m_\tau$ and the Cabbibo angle $V_{us} \sim V_{cd}$.
One then varies the parameters and recalculates these 3 quantities with
the above procedure until agreement is obtained with the experimental
values.  A useful point to note is that whereas
the mass ratios $m_c/m_t, m_\mu/m_\tau$ depend mostly on the parameters
$\rho$ and $ y_I$, which govern respectively the speed of rotation and the
curvature of the rotation trajectory, the Cabbibo angle is sensitive to
$z_I$ which governs the nonplanarity of the trajectory.  The best fit
obtained in \cite{phenodsm} with the central values given by the
Particle Physics Booklet
at that time \cite{databook96}
for $m_c/m_t, m_\mu/m_\tau$ and the Cabbibo angle gives (see,
however, footnote after equation (\ref{x1tilde})):
\begin{equation}
\rho = 4.564, \ \ y_I = 0.0017900, \ \  z_I = 0.0000179.
\label{fitparam}
\end{equation}

With these fitted parameters in hand, one can now calculate the trajectory
to 1-loop level, the result of which is shown in Figure \ref{sphere}.
The speed at which the vector ${\bf r} = (x', y', z')$ rotates with change
in scale $\mu$ is not explicitly shown in this figure but can be gauged
from the locations on the trajectory of the various quark and lepton states
each marked at the scale $\mu$ equal to the mass of that particular state.
The same result will be presented again later in Figure \ref{3Dplot} in
which the $\mu$-dependence is made explicit.  One notices in Figure
\ref{sphere} that the trajectory calculated with the parameters fitted
as above automatically puts the $t$ quark very near the high energy
fixed point $(1,0,0)$, and the neutrinos bunched up near the low
energy fixed point $\frac{1}{\sqrt{3}} (1,1,1)$.  This means that the
qualitative arguments of the last section apply, so that some
reasonable values for the mixing angles and mass ratios can already be
anticipated.  Besides one seee that 
the rotation, as expected, is slow near the high energy
fixed point at $(1, 0, 0)$ so that from $\mu = \infty$ to the scale of
the top mass $m_t$ at 175 GeV, the vector ${\bf r}$ has rotated only
by an angle of about 0.03 radians.  Even down to the scale of the muon mass
$m_\mu$ at 105 MeV, the rotation angle is still of order only 0.3 radians, 
and that
is the reason already mentioned at the beginning of the section why one
expects the 1-loop approximation to be still roughly valid down to
this energy region, with 2-loop contributions presumably
of order of the square
of the 1-loop, leading to about a 30 percent correction.  The rotation,
however, will continue to accelerate as the scale moves further down so that
for scales $\mu$ less than the muon mass, the above 1-loop approximation
will become unreliable.  This means in practice that it should normally be
applied only to the 2 heavier generations of the $U, D, L$ fermion species
and not to neutrinos, except under special circumstances of which there
are some very important examples to be explained later.

Having calculated the rotation trajectory of the vector ${\bf r}$ and
hence of the fermion mass matrix (\ref{mtreewp}), one can now follow the
prescription detailed in the preceding section to evaluate the masses and
state vectors of all the 9 states of the $U, D$ and $ L$ fermion species,
but with the above
proviso that only the results
for the 2 heavier generations are normally to be trusted.  We note, however,
that since in each fermion species the state vectors of the 3 generations
form together an orthonormal triad, the state vector of the lightest
generation is already determined by the state vectors of the 2 heavier
generations which in turn are already determined at the mass scale of the
second heaviest state.  Hence, despite the proviso above, one concludes
that all 3 state vectors can be evaluated with confidence already at the
1-loop level.  This is fortunate, for it means for quarks in particular
that the triads for both $U$ and $D$ quarks are now determined and this
allows one immediately via (\ref{ckmdot}) to evaluate the whole CKM
matrix (apart from the CP-violating phase as explained above).
The result obtained with the parameters in (\ref{fitparam}) is as follows
\cite{phenodsm}:
\begin{eqnarray}
\lefteqn{
\left( \begin{array}{ccc} |V_{ud}| & |V_{us}| & |V_{ub}| \\
                          |V_{cd}| & |V_{cs}| & |V_{cb}| \\
                          |V_{td}| & |V_{ts}| & |V_{tb}|
   \end{array} \right)
   =} \nonumber\\
&& \left( \begin{array}{lll}
   0.9745-0.9762 & 0.217-0.224 & 0.0043-0.0046 \\
   0.217-0.224 & 0.9733-0.9756 & 0.0354-0.0508 \\
   0.0120-0.0157 & 0.0336-0.0486 & 0.9988-0.9994  \end{array} \right),
\label{thckm}
\end{eqnarray}
where the range of values in the entries represent the spread in values
given by the data booklet for the fitted quantities $m_c/m_t, m_\mu/m_\tau$
and the Cabbibo angle.  It is seen that (\ref{thckm}) not only shares the
general features noted in the empirical CKM matrix (\ref{exckm}) as expected
already by the qualitative considerations in the preceding section, but
even agrees quantitatively with the empirical CKM matrix all to within the
quoted experimental errors.

By the same token as for quarks, the above 1-loop calculation for the
rotation trajectory for the vector ${\bf r}$ should give the state vectors
of the 3 charged leptons $\tau, \mu, e$ with some confidence.  Explicitly,
one obtains \cite{phenodsm}:
\begin{eqnarray}
|\tau \rangle & = & (0.996732, 0.076223, 0.026756), \nonumber\\
|\mu \rangle  & = & (-0.075925, 0.774100, 0.628494), \nonumber\\
|e \rangle    & = & (0.027068, -0.628482, 0.777354).
\label{leptonvs}
\end{eqnarray}
However, these by themselves do not allow one to calculate the MNS mixing
matrix, for which one will need also the state
vectors of the  neutrinos.  The present empirical bound on the mass of the
electron neutrino
from e.g.\ tritium decay experiments is of order eV, which in turn
implies that the mass
differences of the heavier neutrinos $\nu_2$ and $\nu_3$ are
restricted by neutrino oscillation experiments to order 0.1 eV or less.
This puts all active neutrinos in the eV mass scale range or lower, which 
is clearly way beyond the range of validity of the above 1-loop calculation.  
Once again, however, there are fortunate special circumstances here that
help us through \cite{bordsm}.  
According to the prescription of the preceding section,
the state vector of the heaviest neutrino $\nu_3$, as for the heaviest 
generation in all other fermion species, is just the vector 
${\bf r}$ taken at the mass scale of $\nu_3$, i.e.\ at eV scale or lower.
According to Figure \ref{sphere}, however, the vector ${\bf r}$ at the 
eV scale would already be very close to the low energy fixed point at 
$\frac{1}{\sqrt{3}}(1, 1, 1)$ and can be well approximated by it, thus:
\begin{equation}
|\nu_3 \rangle \sim \frac{1}{\sqrt{3}}(1, 1, 1).
\label{nu3v}
\end{equation}
So applying the formula (\ref{ckmdot}) for mixing matrix elements
with the help of (\ref{leptonvs}) above gives immediately:
\begin{eqnarray}
U_{\mu 3} = \langle \mu|\nu_3 \rangle & = & 0.7660, \nonumber\\
U_{e 3}   = \langle e  |\nu_3 \rangle & = & 0.1016.
\label{Umu3e3}
\end{eqnarray}
Again, one sees that these results agree with present data 
(\ref{exmns}) within the experimental errors.  That the mixing element
$U_{\mu 3}$ should be large and $U_{e 3}$ small was expected already from
the qualitative considerations of the preceding section with elementary
differential geometry.  That $U_{\mu 3}$ should turn out, however, to be
near maximal in agreement with oscillation experiments on atmospheric
neutrinos \cite{SuperK,Soudan}, and that the CHOOZ angle to be within the
experimental bounds \cite{Chooz}, are particular achievements of the 1-loop
calculation above.

The mass of $\nu_2$, being by definition even lower than that of $\nu_3$,
will be even nearer the low energy fixed point $\frac{1}{\sqrt{3}}(1,1,1)$.
However, one cannot as yet determine the state vector of
$\nu_2$, which being essentially the tangent vector to the trajectory of
${\bf r}$ at the fixed point, will depend more on the
actual trajectory,
namely on how
the trajectory approaches the fixed point.  It is thus not expected to be
well reproduced by the above 1-loop calculation which will be
unreliable at these energies.  Indeed, should one persist nevertheless
to calculate $|\nu_2 \rangle$ and hence the mixing element $U_{e2}$ as
we did in \cite{phenodsm} before we had realized clearly the limitations
of the 1-loop approximation, one obtains $U_{e2} = 0.24$ which is largish
as solar neutrino experiments show it to be but lies outside present
experimental limits.  We shall return later to comment further on this
point\footnote{The above treatment of neutrino oscillations to 1-loop
level in
DSM \cite{bordsm}
updates and supercedes our earlier treatment in \cite{nuosc}.  The
older treatment made additional assumptions on neutrino masses, and applied
to only the vacuum oscillation solution for solar neutrinos.  The present
treatment needs no special assumptions for neutrinos and applies as well
to the experimentally favoured large mixing angle (LMA) solution for
solar neutrinos, besides accounting properly for the limitations of the
1-loop approximation.}.

Having now exhausted the consequences of the above 1-loop calculation on
the mixing matrices, let us turn next to those on the mass ratios between
generations.  Here the result is much less conclusive, for several
reasons.
First, in contrast to the above calculation of the mixing parameters,
the calculation of lower generation masses depends on the assumption
that the normalization $m_T$ of the mass matrix written in the form
(\ref{mfact}) is roughly constant with changing scale, which may be
reasonable over small scale changes, such as that between the 2
heavier generations considered so far, but would be unreliable when
larger scale changes are involved as when the lightest states are also
taken into account.   Secondly,
of the 3 measured mass ratios involving only the 2 heavier generations,
2 ratios ($m_c/m_t,m_\mu/m_\tau$) have already been used to fit the
parameters
of the model, leaving only $m_s/m_b$ which is poorly measured.  Although a
value for $m_s/m_b = 0.039$ is obtained which is within the wide experimental
bounds, no great significance can be given to the agreement.  Thirdly, the
remaining lightest members of the $U, D, L$ species, namely $u, d$ and $e$,
all lie outside the scale range of applicability of the 1-loop calculation
so that its predictions for their masses cannot be trusted.  Should  one
persist as we did in \cite{phenodsm}, one would obtain for $m_e$ a value of
6 MeV, an order of magnitude off the empirical value 0.5 MeV, which is
already not
too bad, considering that it is an extrapolation in a logarithmic
scale over several orders of magnitude.  Again, we shall return later for
a comment on this.  Fourthly, for the remaining $u$ and $d$ quarks, these
have the additional complication of being tightly confined, while
the prescription given in the last section for
calculating fermion masses applies only to free particles.
Indeed, we do not know at present how
to calculate the masses of these tightly confined states.  The masses of
$u$ and $d$ quoted in the data booklets \cite{databook96,databook} were
determined at values of the running scale of 1 and 2 GeV respectively.
If we were to
define the masses of $u$ and $d$ as the leakage of the vector ${\bf r}$
at these scales to respectively the $u$ and $d$ states which we already
know, we would
obtain masses of the order of MeV, which is of the right order
of magnitude.  But we are not at all confident that this is the correct
prescription.  As for neutrino masses, one has no predictions so far from
the above considerations, except that they have in general nonzero
masses. 

To summarize, with 3 parameters fitted to experiment, one has calculated
to 1-loop approximation the rotation trajectory which allows one then to
determine the mixing matrices and mass ratios between generations.  Having
now explored all possibilities, one finds agreement with data to within
experimental errors for all quantities which are inside the estimated range
of applicability of the 1-loop approximation.  These include all 9 elements
of the CKM matrix $|V_{rs}|, r = u, c, t,\, s = d, s, b$, the 2 elements
$|U_{\mu 3}|, |U_{e 3}|$ of the MNS matrix, as well as the 3 mass ratios
$m_c/m_t, m_s/m_b, m_\mu/m_\tau$, and together account for 8 of the 25 or
so independent ``fundamental'' parameters of the Standard Model as usually
formulated.  For the remaining quantities which lie outside the range of
applicability of the 1-loop approximation, namely $|U_{e 2}|, m_e, m_u, m_d$,
if one persists nevertheless with the 1-loop approximation, one obtains
sensible values of roughly the right order as expected already from our
previous qualitative considerations but lying outside experimental bounds.
In other words, apart from the one important piece of the puzzle of
CP-violation of which one has still found no trace, the DSM scheme taken
to the 1-loop level seems at present to be in the happy position of having
been shown to be right in all cases where it is expected to be right and
to have only qualitative but not
quantitative agreement with data
in circumstances where the 1-loop
approximation made is expected to be unreliable \cite{bordsm}.

The above result has one perhaps unexpected aspect in that the rotation
effect crucial for its derivation is obtained from 1-loop diagrams with
heavy dual colour Higgs boson exchanged where normally one would expect
that at the low energies one is dealing with the heavy bosons could be
integrated out and largely ignored.  We think, however, that there are
special circumstances here which differ from the normal expectation.
Given the initial assumption that these bosons exist, then they will
in any case contribute at 1-loop to the renormalization of the fermion 
mass matrix as calculated.  The terms proportional to $\ln \mu$ which 
affect the rotation occur as wave function renormalization and are not 
among those shown by Appelqvist and Carrazone to be of order $s/M^2$ 
in their decoupling theorem \cite{decoupth}.  Nevertheless, one might 
have expected them to be overwhelmed at low energy scales both by the 
higher loop contributions of these heavy bosons, and by the loop 
diagrams of the Standard Model particles, such as gluons and electroweak 
gauge and Higgs bosons.  However, as already mentioned, in the special 
case of mass matrix rotation that we are looking at, the higher loop 
corrections due to dual colour bosons are small because of the proximity 
to the rotational fixed points in the scale regions under consideration, 
while, even more importantly, 
the contributions of the Standard Model particles to the rotation give
zero.  For this reason, it appears that for lack of competition, the 1-loop
contribution of the heavy bosons will still dominate and give already
a reasonable approximation.  However, without a more detailed investigation,
one cannot go any further than this qualitative observation, and can at
present only leave the positive results to speak for themselves.

The conclusion of general agreement with data, however, holds at this
moment only as regards the fermion mass and mixing patterns, which
are the only pieces of data
so far explored.  Consequences in other areas have yet to be
examined later in section 7.  Besides, the success of the predictions,
of course, need by no means imply that the whole chain of arguments
leading to the predictions are correct.  Our next task therefore is to
examine which of the arguments are essential and which are not for obtaining
the above positive result.

\setcounter{equation}{0}

\section{Direct Empirical Support for Mass Matrix Rotation}

The DSM 1-loop calculation reported above giving good agreement with
experiment on fermion mass and mixing parameters was first performed
numerically \cite{ckm,phenodsm} but has since been checked by analytic
calculations under certain approximations \cite{bordsm}, and being backed
up further by the qualitative considerations of section 4, seems unlikely
to be pure coincidence.  What is unclear, however, is whether the
apparent success can be ascribed to the assumptions that have been made,
and if so to what extent and to which of them.  In other words, we wish to
ask what the above
calculation has actually taught us about the underlying physics.

It has already been pointed out in \cite{phenodsm} that although the concept
of nonabelian duality as described in section 2 and the identification of
the dual colour symmetry with the horizontal symmtery of generations are
seminal in first of all offering a geometric explanation for the existence
of 3 fermion generations, and secondly in suggesting a new framework for
calculating the fermion mass hierarchy and mixing phenomena, they are not
absolutely essential for obtaining the desired result.  Indeed, neither the
Higgs potential (\ref{Vofphi}) nor the Yukawa coupling term (\ref{Yukawa})
from which the calculation develops have been shown to follow logically
from nonabelian duality, and so long as one has a horizontal symmetry
with these two ingredients, the same calculation can be carried through
with the same apparent success without any reference to nonabelian duality
or to its identification with generations.  Furthermore, as far as the
qualitative features of the fermion mass hierarchy and  mixing are
concerned, it would appear from the discussion of section 4 that what is
really crucial is that the mass matrix should rotate and that there should
be rotational fixed points at infinite and zero energy scales.  Where the
suggested Higgs potential and Yukawa coupling come in is really only in
supplying the detailed fit to the experimental numbers.  Superficially at
least, it would seem conceivable that given a rotation trajectory
depending on several parameters, so long as it has the same rotational fixed
points as above, then very similar results would obtain, regardless
of the theoretical premises from which the rotation trajectory may arise.

One can go even further and ask whether the rotation itself is necessary.
To answer this question, one can proceed as follows, namely by turning the
problem around and discarding at first even the assumption of a rotational
trajectory but seeking instead 
evidence for it directly from experimental data.  
This seems at first sight a tall order, but turns out actually to be
practicable under certain assumptions as we shall now explain.  From the
discussion in section 4, one sees that so long as the mass matrix, for
whatever reason, can have different orientations at different energy scales,
then the usual definition of fermion masses and state vectors will have already
to be refined.  In particular, even a rank 1 mass matrix (i.e.\ with only one
nonzero eigenvalue) will acquire nonzero masses for the 2 lower generations,
and even when the mass matrices of up and down fermions are aligned in
orientation at the same scale, there will be nontrivial mixing between them
when the difference in orientation at different scales is taken into
account.  And these effects are immediately calculable once the difference
in orientation of the mass matrix at different scales is known.  Suppose
then we assume that all masses for the 2 lower generations as well as the
mixing between up and down states arise only from this effect, we can then
turn the argument around and ask what differences in orientations are
necessary at different scales to produce the experimentally observed mass
ratios and mixings.  When this information is extracted from the data and
plotted as a function of the scale, then if the hypothesis of a rotational
trajectory is indeed correct, the data points will not be randomly
scattered but 
should all lie on some smooth
curve.  And if they do, one will then have 
evidence for the rotational trajectory
directly from experimental data.

\begin{figure}
\centering
\hspace*{-3.5cm}
\input{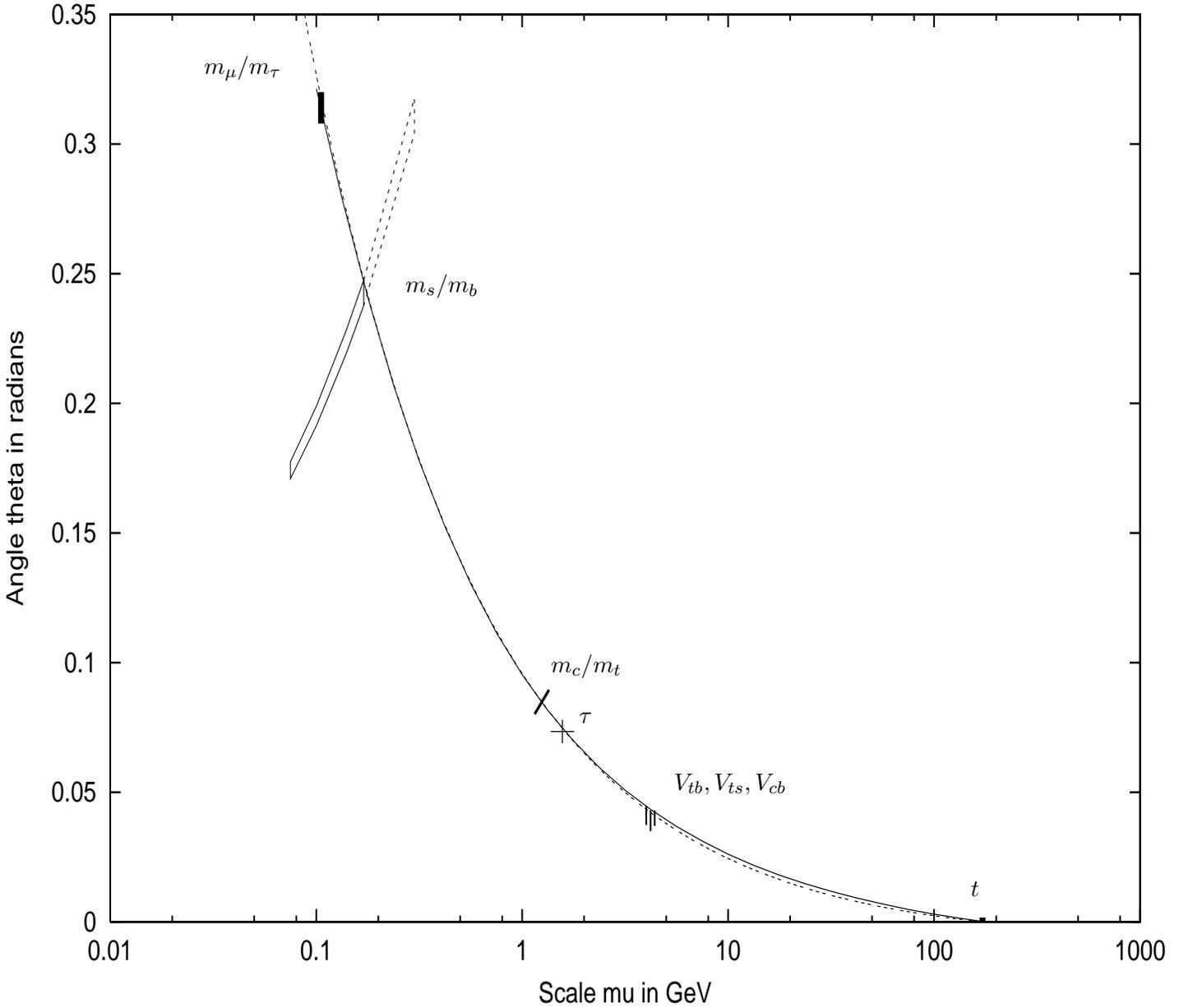}
\caption{The rotation angle changing with scale as extracted from data on
mass ratios and mixing angles (in the ``planar'' approximation with only 
the 2 heavier generations) and compared with the best fit to the data
with an exponential (dashed curve) and the earlier calculation by DSM
(full curve) \cite{phenodsm}.}
\label{planero}
\end{figure}

To see practically how this can be done, let us first work
in the simplified scenario with only the 2 heavier generations in each
fermion species, which will bring
out many of the salient points in a transparent manner.  Besides, it
will be seen to be already a good approximation for the high mass
scale region.
The problem now being planar, the difference in
orientation of a vector between 2 scales is given by an angle 
which is additive in the
sense that the difference $\theta_{13}$ from scale 1 to scale 3 equals
the sum $\theta_{12} + \theta_{23}$ of the difference from scale 1 to
scale 2 and that from scale 2 to scale 3.  As explained above, we start
with a rank 1 mass matrix (in the hermitian Weinberg notation), which is
thus necessarily of the factorizable form (\ref{mfact}) given in terms of
a single vector ${\bf r}$ and it is the dependence of this vector on the
energy scale we wish to trace, using experimental data on mass ratios
and mixing matrix elements.  Consider first the mass ratio $m_c/m_t$ with
$m_t = 174.3 \pm 5.1 \ {\rm GeV}$ and $m_c = 1.15 - 1.35 \ {\rm GeV}$ as given
in \cite{databook}.  By (\ref{planeleak}) this 
is just $\sin^2 \theta_{tc}$, from which 
one easily extracts the value of $\theta_{tc}$ together with
the appropriate errors.  Similarly, using (\ref{planemix}) one extracts
again easily from the value of the CKM matrix element $|V_{tb}|$ the
value of $\theta_{tb}$.
The same can be done for all other pieces of data involving the 2 heavier
generations in the $U, D, L$ fermion species to deduce the differences in
orientation between the different mass scales. 
The result \cite{cevidsm} is plotted in Figure \ref{planero}, where one has
made use of additivity to deduce, for example, that $\theta_{ts} =
\theta_{tb}
+ \theta_{bs}$, with $\theta_{tb}$ already obtained from
$|V_{tb}|$ above, and $\theta_{bs}$ from the
mass ratio $m_s/m_b$ \cite{databook96,databook}.
One sees in the figure that the extracted data points all lie comfortably
on a smooth curve, which can thus be regarded as empirical evidence for
rotation, i.e. for the vector ${\bf r}$ tracing out a trajectory as the scale
changes.  Indeed, the trajectory traced out by the data is surprisingly close 
to that calculated in \cite{phenodsm} several years before the data were
examined in this way.  A best fit with MINUIT to the data points in
Figure \ref{planero} gives the dotted curve shown which is seen to be
hardly distinguishable from the full curve obtained from the calculation
in \cite{phenodsm}.

The evidence cited above for the rotation hypothesis based on Figure
\ref{planero} is subject to the planar approximation which takes into
account only the 2 heavier generations in each fermion species.  However,
it can be seen that for the numbers so far extracted the approximation is
already sufficiently accurate.  Take for example the angle $\theta_{ts}$
which was extracted above using additivity based on the assumption that
the 3 vectors ${\bf r}$ at the 3 mass scales of $t, b$ and $s$ all lie
on the same plane, which is of course not exact.  Indeed, the
deviation from planarity is given by the Cabbibo angle, i.e.\ the angle
between the state vectors of the $u$ and $d$ quarks which are respectively
normal to the planes spanned by the state vectors of $t$ and $c$ and those
of $b$ and $s$.  From the empirical value of the Cabbibo angle of around
0.22 radians and the fact that all the angles exhibited in the Figure
\ref{planero} depend on the square of the Cabbibo angle, one estimates
that the error committed by the planar approximation is of order of only 4
percent and hence does not affect the significance of the evidence for
rotation claimed above.

The planar approximation, however, worsens as the scale lowers further
and cannot therefore be used to extend the analyses to the electron and
neutrinos regions which are of crucial physical interest.  Besides, it
does not reveal the details in the off-planar direction which could in
principle contain surprises upsetting the above positive result.  For this
reason, a full 3 generation repeat of the above analysis is necessary in
order to draw a firm conclusion.  Since the mass matrix is of rank 1 by
our initial assumption, it is still factorizable in terms of a single
vector ${\bf r}$ which changes orientation with changing scale.  And to
extract the variations of this vector between different mass scales is no
different in principle for 3 generations than for 2, only in practice more
complicated.  There are more angles involved and simple additivity no
longer applies, but with patience the analysis can be carried through.  
Take, for example, the interesting case of the vector ${\bf r}$ taken at
the mass scale of the heaviest neutrino $\nu_3$.  The MNS mixing matrix
elements $|U_{\mu 3}|$ and $|U_{e 3}|$ are now constrained by
oscillation experiments with respectively atmospheric \cite{SuperK,Soudan}
and terrestrial neutrinos \cite{Chooz,K2K} to within the following bounds:
$1/3 < |U_{\mu 3}|^2 < 2/3$ and $|U_{e 3}|^2 < 0.027$.  Now (\ref{ckmdot})
gives $|U_{\mu 3}| = \langle \mu|\nu_3 \rangle$ and 
$|U_{e 3}| = \langle e|\nu_3 \rangle$.
Hence, if the state vectors $|\mu \rangle, |e \rangle$ are exactly known as
well as the quantities $|U_{\mu 3}|$, $|U_{e 3}|$, then 
the state vector $|\nu_3 \rangle$ 
is determined up to discrete sign ambiguities.  Even as matters stand, 
where the quantities involved are known only 
within certain experimental bounds, it still means that the state vector 
of $\nu_3$ will be constrained within well defined limits, which can then
be checked for consistency with the rotation hypothesis.  Furthermore, we
recall from our earlier discussion that $|\nu_3 \rangle$ is supposed to
have almost reached the asymptotic limit of the fixed point at zero scale 
predicted by the DSM, which prediction could also thus be directly
confronted with the limits extracted from data.

In any case, the full 3 generation analysis has been carried out tracing
the trajectory of ${\bf r}$ over some 14 orders of magnitude in energy
scale from the mass scale of the top quark to that of the second 
heaviest neutrino
$\nu_2$ with the result shown in Figure \ref{3Dplot}.  The technical details
involved can be found in \cite{cevidsm}.  Figure \ref{3Dplot} gives in a
3-dimensional plot the second 
and third components of the vector ${\bf r}$
extracted from the data for various scales corresponding to the masses
of the fermions states.  For technical reasons, these are given in a frame
defined by the $U$ quark triad as frame vectors (i.e. ${\bf v}_t = (1,0,0),
{\bf v}_c = (0,1,0), {\bf v}_u = (0,0,1)$), not in the old frame where the 
high energy fixed point appears as $(1,0,0)$.  Apart from the information 
from the masses of $u$ and $d$ which is ambiguous for reasons already 
explained and that from the solar neutrino angle which cannot easily be 
presented in this figure (see later, however), the data points shown 
represent all the information on the vector ${\bf r}$
that could at present be extracted from 
experiment on fermion masses and mixing.
And it can be seen in the figure that all these data, instead
of being a random collection of points, 
are perfectly consistent with them lying on a smooth 3-D curve.  
The consistency 
can be scrutinised further in the 3 projections of Figure \ref{3Dplot} on 
to the 3 coordinate planes shown in Figures \ref{etazeta}, \ref{mueta}, 
and \ref{muzeta}, where it is seen in Figure \ref{etazeta} that even the 
oscillation data from solar neutrinos missed out in Figure \ref{3Dplot}
satisfy this consistency, as indicated there by the dotted curve.  This 
overall consistency with the rotation hypothesis seems quite nontrivial 
especially in the high scale region above the muon mass, given the accuracy 
of the data there.

\begin{figure*}
\vspace*{-3cm}
\centering
\includegraphics[scale=0.9]{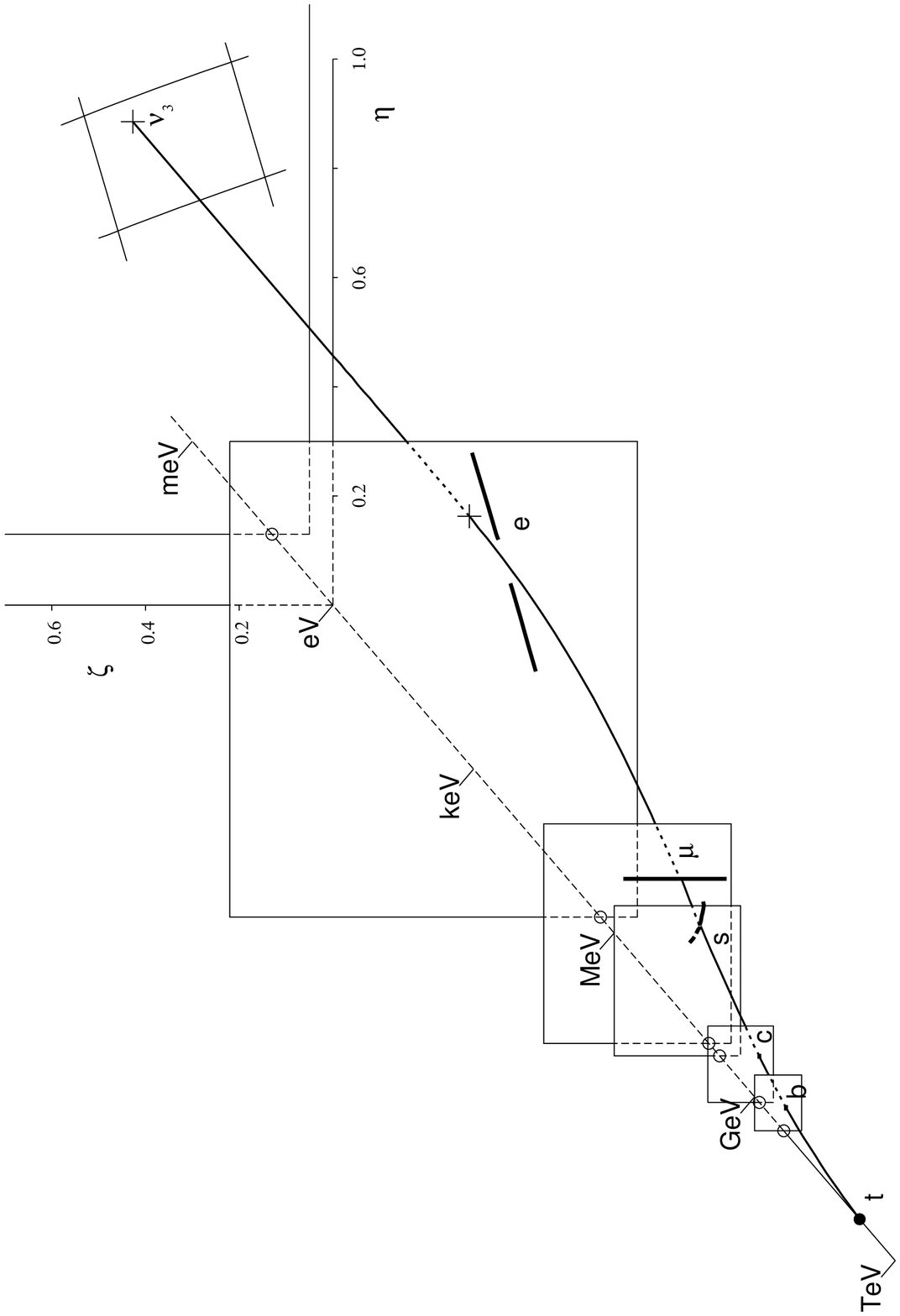}
\end{figure*}

\begin{figure}
\vspace{8cm}
\caption{A plot of the rotating vector ${\bf r}(\mu)$ as extracted from
existing data on fermion mass ratios and mixing parameters, where its 2nd
and 3rd components $\eta$ and $\zeta$ are plotted as
functions of $\ln \mu$, $\mu$ being the energy scale.  The experimentally
allowed values at any one scale are represented as an allowed region on a
plaquette, with the scale corresponding to a plaquette being given by the
intersection, denoted by a small circle, of its left-most boundardy with
the $\mu$-axis.  For example, the first small plaquette on the left of the
figure corresponds to the scale $\mu = m_b$, on which plaquette the allowed
region for ${\bf r}(\mu) = {\bf v}_b$ is very small because of the small
experimental error on the CKM matrix elements $V_{tb}, V_{cb}$ and $V_{ub}$.
The last plaquettte on the right, on the other hand, corresponds to the
scale $\mu = m_{\nu_3}$, on which plaquette the allowed region for
${\bf r}(\mu)$ is a rough rectangular area bounded by the data on $\nu$
oscillations from atmospheric neutrinos and from the Chooz experiment.  The
curve represents the result of a DSM one-loop calculation from an earlier
paper \cite{phenodsm} which is seen to pass through the allowed region on
every plaquette except that for the electron $e$.  For further explanation
of details, please see text.}
\label{3Dplot}
\end{figure}

\begin{figure}
\vspace*{-3cm}
\centering
\includegraphics[scale=0.8]{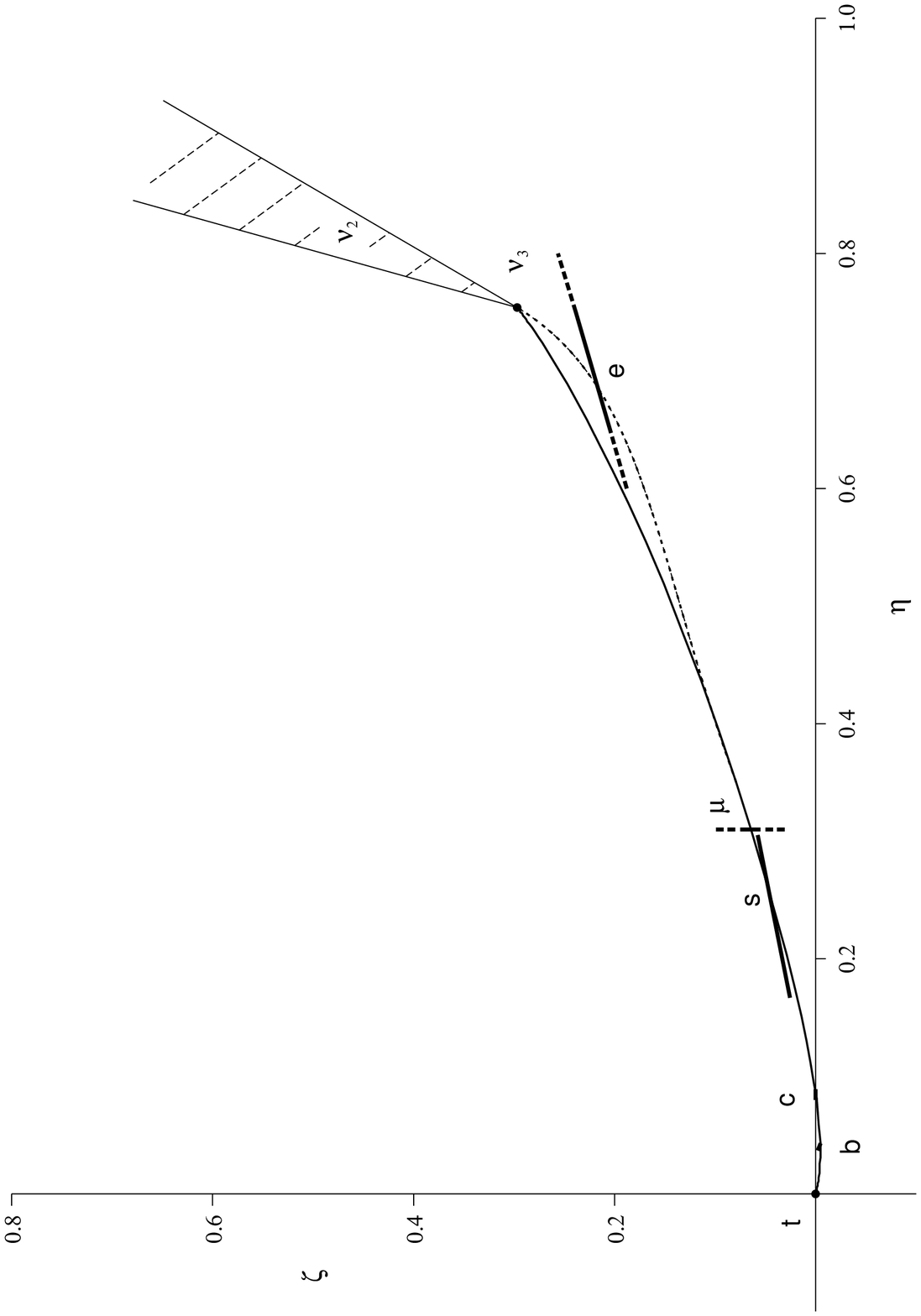}
\caption{Projection of Figure \ref{3Dplot} on to the $\eta\zeta$-plane.
The full curve represents the DSM one-loop calculation of \cite{phenodsm}
and the dashed curve its suggested deformation at low scales to fit the
data on $m_e$ and $U_{e2}$.}
\label{etazeta}
\end{figure}

\begin{figure}
\vspace*{-3cm}
\centering
\includegraphics[scale=0.8]{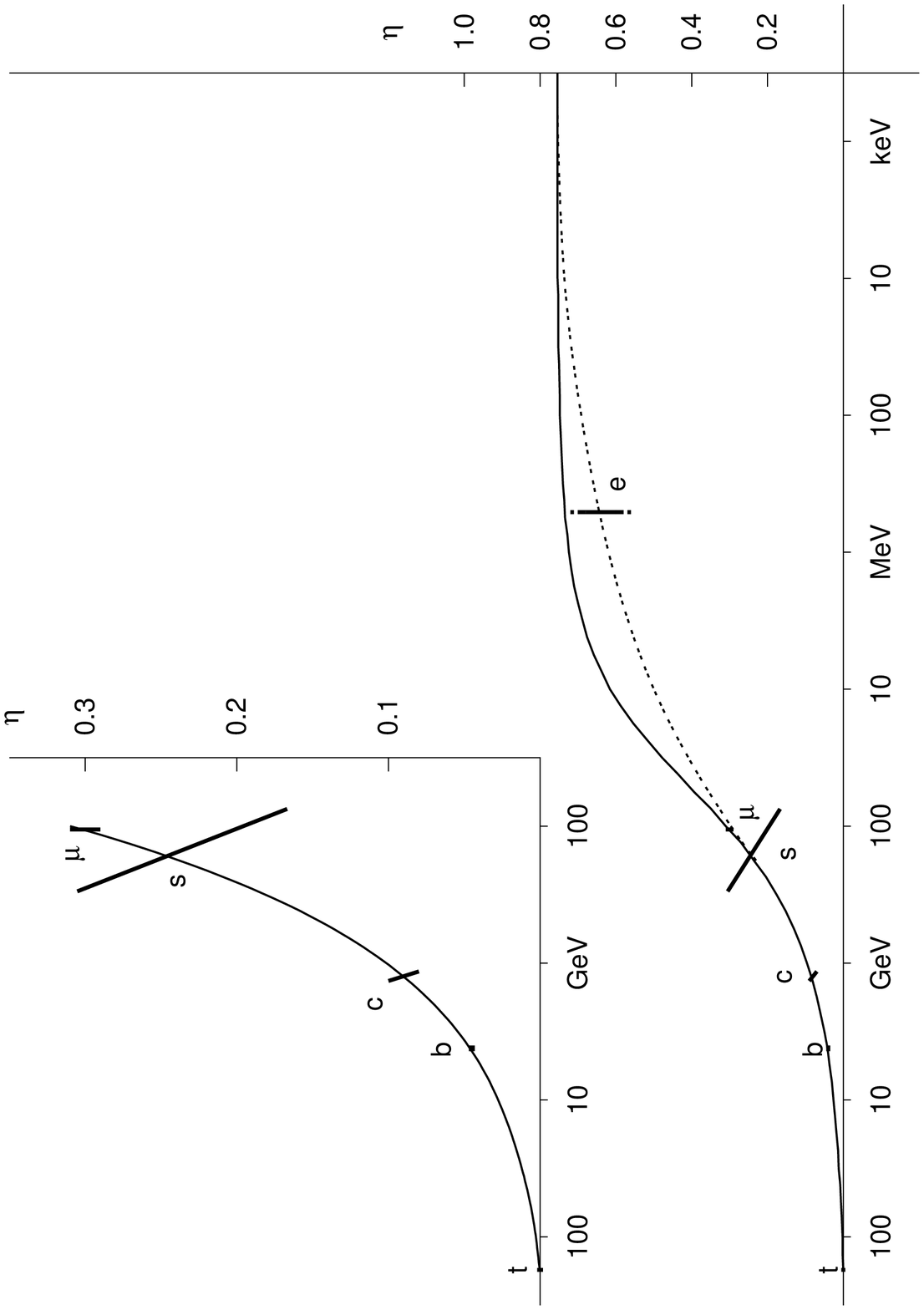}
\caption{Projection of Figure \ref{3Dplot} on to the $\mu\eta$-plane.
The full curve represents the DSM one-loop calculation of \cite{phenodsm}
and the dashed curve its suggested deformation at low scales to fit the
data on $m_e$ and $U_{e2}$.}
\label{mueta}
\end{figure}

\begin{figure}
\vspace*{-3cm}
\centering
\includegraphics[scale=0.8]{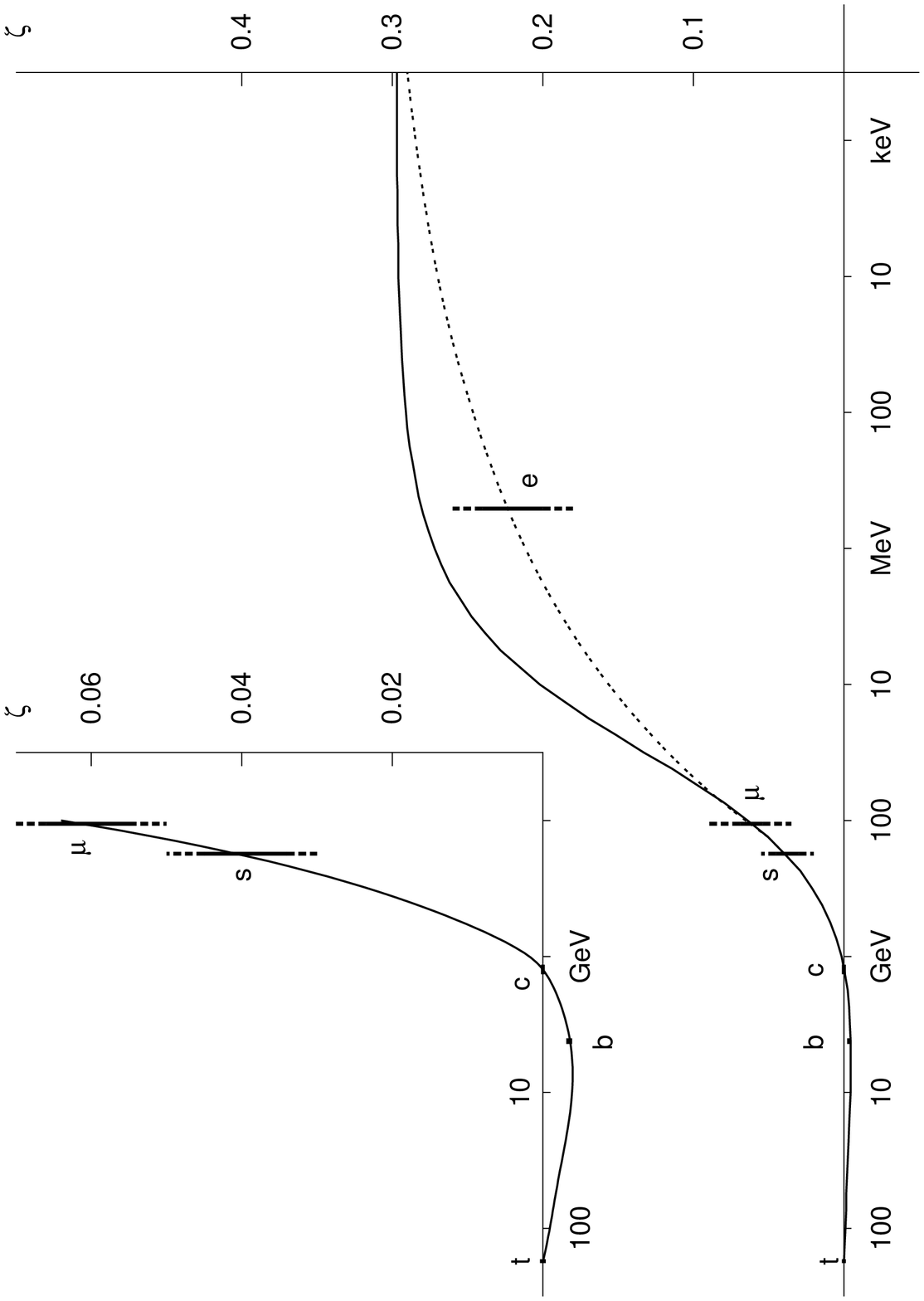}
\caption{Projection of Figure \ref{3Dplot} on to the $\mu\zeta$-plane.
The full curve represents the DSM one-loop calculation of \cite{phenodsm}
and the dashed curve its suggested deformation at low scales to fit the
data on $m_e$ and $U_{e2}$.}
\label{muzeta}
\end{figure}

The full curve in the Figure \ref{3Dplot} and its 3 projections represents
the 1-loop result from the DSM calculation of \cite{phenodsm} described
in section 5.  It is seen that above the scale $\mu = m_\mu$ it passes 
through all the data points within errors.  That this is the case was
already discussed 
in the preceding section, but as displayed in these figures
together with the experimental errors, it is easier to appreciate the
significance of the surprisingly good agreement between the calculation
and experiment.  At scales below the muon mass, the 1-loop curve deviates
from the data as expected, thus missing the 2 allowed regions for ${\bf r}$
deduced respectively from the electron mass and the solar neutrino angle
$U_{e2}$.  It would be interesting to enquire, as we are now trying to do,
whether a 2-loop calculation would improve the agreement in the low scale
range.

Although the DSM scheme has done extremely  well in fitting the data with 
only 3 parameters, one can still wonder whether all of its details are
strictly necessary. 
Given that the vector ${\bf r}$ extracted directly from the data already 
seems to trace out a rotation trajectory, it would appear that, 
as described in section 4, if one used a 
rotating mass matrix of rank 1 having
fixed points at infinite and zero scales, 
then with several adjustable parameters to fit the rotational 
trajectory, one could probably already manage quite well phenomenologically 
without appealing to the other details.  We take this to mean that minor 
modifications of the DSM scheme as it now stands, which might in future 
be found necessary for theoretical reasons, could well leave intact the 
phenomenological successes so far achieved.  What we have in mind is the 
possibility that a closer examination of duality may lead to some unique 
forms for the Higgs potential and Yukawa coupling possibly differing 
slightly from those at present assumed but yet achieving similar 
phenomenological success.  If so, that would be ideal.

\setcounter{equation}{0}

\section{Other DSM Consequences}

As seen above, the Dualized Standard Model has been quite successful in
explaining fermion mass hierarchy and mixing, although one is not entirely
certain as yet how much of the details in its structure is
essential for this success.  There is, however, still another angle to
explore before one can properly gauge the significance of this 
seeming success.
The DSM involves new assumptions beyond those already tested by experiment
within the context of the conventional Standard Model, and is therefore
bound to give some new physical predictions.  One has thus to ask first,
whether these new predictions agree with all existing experiment, and
secondly, if they manage to survive these tests, whether they can be further
tested by experiment in the near future.

\begin{figure}
\vspace*{-2cm}
{\unitlength=1.0 pt \SetScale{1.0} \SetWidth{1.0}
\begin{picture}(200,200)(0,0)
\Photon(65,28)(65,90){3}{5.5}
\Line(65,32)(25,12)
\Line(65,28)(25,8)
\Line(65,32)(100,12)
\Line(65,28)(100,8)
\Line(65,90)(25,110)
\Line(65,90)(100,110)

\Text(10,10)[]{$N$}
\Text(115,10)[]{$N$}
\Text(80,60)[]{$X$}
\Text(10,113)[]{$\mu$}
\Text(115,113)[]{$e$}
\Text(65,-5)[]{$(a)$}

\Line(300,60)(330,30)
\Line(300,60)(330,90)
\Photon(300,60)(260,60){3}{3.5}
\Line(260,63)(220,63)
\Line(260,57)(220,57)
\Line(260,63)(260,57)
\Text(345,93)[]{$\mu$}
\Text(345,27)[]{$e$}
\Text(285,75)[]{$X$}
\Text(210,52)[]{$s$}
\Text(210,68)[]{$u$}
\Text(265,-5)[]{$(b)$}

\end{picture} }
\vspace*{7mm}
\caption{Diagrams representing schematically (a) $\mu-e$ conversion in
nuclei, (b)
$K_L \rightarrow \mu e$
decay, as FCNC effects via the exchange
of heavy bosons $X$ carrying generation index.}
\label{fcnceg}
\end{figure}
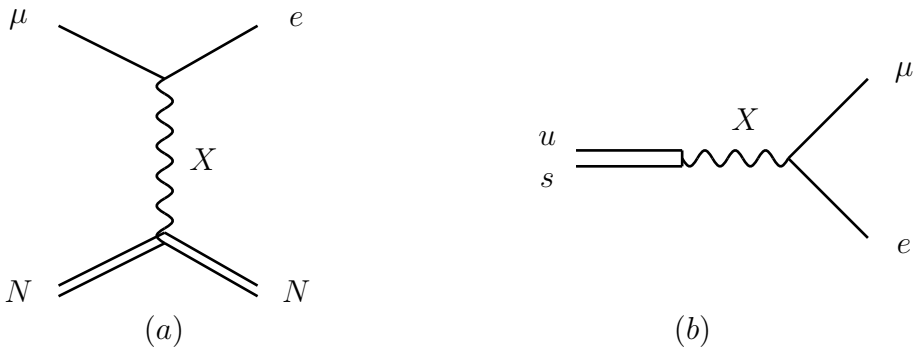

One obvious direction to explore is flavour-violation which can occur
in the DSM scheme in two ways.  The first
type is in common with all horizontal symmetry models in which the
horizontal symmetry is mediated by bosons carrying the generation index.
The exchange of such bosons would lead to flavour-changing neutral current
(FCNC) effects, of which $K$ meson decay to $\mu e$ and $\mu-e$ conversion
in nuclei are typical examples, as illustrated by the diagrams in Figure
\ref{fcnceg}.  The masses $M_X$ of the mediating bosons are presumably high
or otherwise they should have been seen already, and they are not.  If so,
then at the low energies where FCNC effects are studied in experiment,
the reaction amplitudes would be suppressed by factors of order $s/M_X^2$,
leading to suppression in rates of order $(s/M_X^2)^2$.  Hence, predicted
rates of flavour-violations of this type can always be made sufficiently
small to satisfy whatever experimental bounds by making $M_X$ large,
so long as no upper bound for $M_X$ is prescribed by the theory.  For
this reason, for flavour-violating effects of this type, present bounds
from experiment pose no imminent threat to the DSM scheme, nor indeed to
any horizontal symmetry model.

Special to the DSM, however, is another type of flavour-violations
which is potentially much more dangerous.  One crucial property of the
DSM explanation of fermion mass hierarachy and mixing
is the rotation of the mass matrix with changing scales.  And mass
matrix rotation means that a mass matrix diagonal at one scale will in
general no longer be diagonal at another scale.  For example, suppose we
examine Compton scattering of a photon from an electron and the lepton
mass matrix rotates as suggested in DSM.  Then as detailed in section 4,
the mass matrix is diagonal at the mass scales of the leptons, but not in
general at other scales, and in particular not at the energy scale at which
the Compton scattering experiment is performed.  The reaction amplitude,
which depends on the mass matrix, can thus be expected also not to be
diagonal in the flavour states, hence leading to nonzero cross sections
for the nondiagonal, flavour-violating reaction:
\begin{equation}
\gamma e \rightarrow \gamma \tau,
\label{comptone}
\end{equation}
for example.  In other words, purely kinematically, one could expect
flavour-violation to result by virtue alone of the rotating mass
matrix.  And in a scheme such as DSM where the rotation speed, being
tied to the fermion mass and mixing patterns and not adjustable to satisfy
experimental bounds on flavour-violating reactions, the confrontation is
potentially very dangerous.

This new type of flavour-violation due to a rotating mass matrix, to which
we have given the name ``transmutation'' for easy reference, we have studied
in some detail.  It was found that, with the rotation speed constrained by
the fermion mass and mixing pattern, transmutation effects from kinematics
alone can be appreciable.  For example, for the reaction:
\begin{equation}
e^+ e^- \longrightarrow \mu^+ \tau^-,
\label{bhabha}
\end{equation}
the integrated cross section estimated from a rotation speed read from,
for example, Figure \ref{planero} or \ref{3Dplot}, is about 80 fb
\cite{transbhar} at $\sqrt{s} = 10.85$ GeV, at which energy very high
statistics is being collected by experiments such as BaBar \cite{Babar}
and Belle \cite{Belle}.  Although by itself this does not seem a large cross
section, in view of the sensitivity of the above modern experiments with
integrated luminosity of order 100 fb$^{-1}$, it is in fact frighteningly
large, and should in principle be already detectable.

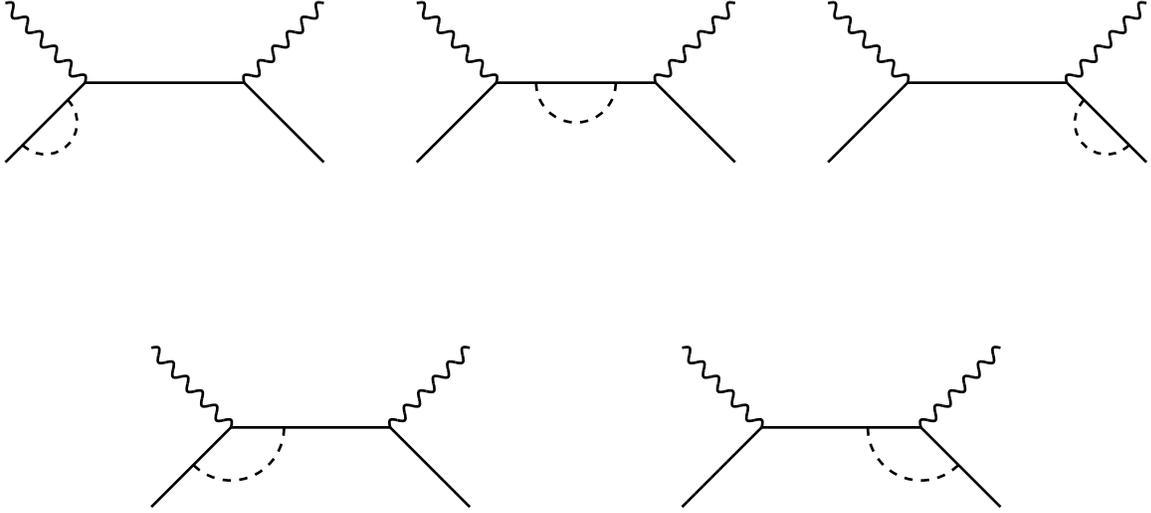
\begin{figure}[ht]
\begin{center}
{\unitlength=1.0 pt \SetScale{1.0} \SetWidth{1.0}
\begin{picture}(350,250)(0,0)

\Line(35,50)(95,50)
\Photon(35,50)(5,80){-2}{5.5}
\Photon(95,50)(125,80){2}{5.5}
\Line(5,20)(35,50)
\Line(95,50)(125,20)
\DashCArc(35,50)(20,225,0){3}

\Line(235,50)(295,50)
\Photon(235,50)(205,80){-2}{5.5}
\Photon(295,50)(325,80){2}{5.5}
\Line(205,20)(235,50)
\Line(295,50)(325,20)
\DashCArc(295,50)(20,180,315){3}

\Line(135,180)(195,180)
\Photon(135,180)(105,210){-2}{5.5}
\Photon(195,180)(225,210){2}{5.5}
\Line(105,150)(135,180)
\Line(195,180)(225,150)
\DashCArc(165,180)(15,180,0){3}

\Line(-20,180)(40,180)
\Photon(-20,180)(-50,210){-2}{5.5}
\Photon(40,180)(70,210){2}{5.5}
\Line(-50,150)(-20,180)
\Line(40,180)(70,150)
\DashCArc(-35,165)(12,225,45){3}

\Line(290,180)(350,180)
\Photon(290,180)(260,210){-2}{5.5}
\Photon(350,180)(380,210){2}{5.5}
\Line(260,150)(290,180)
\Line(350,180)(380,150)
\DashCArc(365,165)(12,135,315){3}

\end{picture}}
\end{center}
\vspace*{5mm}
\caption{1-loop diagrams contributing to transmutation in Compton scattering.}
\label{compton1l}
\end{figure}

\begin{figure}[ht]
\begin{center}
{\unitlength=1.0 pt \SetScale{1.0} \SetWidth{1.0}
\begin{picture}(350,250)(0,0)

\Photon(165,25)(165,75){2}{6.5}
\Line(165,75)(130,100)
\Line(130,0)(165,25)
\Line(165,25)(200,0)
\Line(200,100)(165,75)
\DashCArc(165,25)(20,215.5,325.5){3}

\Photon(30,25)(30,75){2}{6.5}
\Line(30,75)(-5,100)
\Line(-5,0)(30,25)
\Line(30,25)(65,0)
\Line(65,100)(30,75)
\DashCArc(12.5,12.5)(12,215.5,35.5){3}

\Photon(300,25)(300,75){2}{6.5}
\Line(300,75)(270,100)
\Line(270,0)(300,25)
\Line(300,25)(335,0)
\Line(335,100)(300,75)
\DashCArc(317.5,12.5)(12,145.5,325.5){3}

\Photon(165,175)(165,225){2}{6.5}
\Line(165,225)(130,250)
\Line(130,150)(165,175)
\Line(165,175)(200,150)
\Line(200,250)(165,225)
\DashCArc(165,225)(20,35.5,144.5){3}

\Photon(30,175)(30,225){2}{6.5}
\Line(30,225)(-5,250)
\Line(-5,150)(30,175)
\Line(30,175)(65,150)
\Line(65,250)(30,225)
\DashCArc(12.5,237.5)(12,325.5,145.5){3}

\Photon(300,175)(300,225){2}{6.5}
\Line(300,225)(270,250)
\Line(270,150)(300,175)
\Line(300,175)(335,150)
\Line(335,250)(300,225)
\DashCArc(317.5,237.5)(12,35.5,215.5){3}

\end{picture}}
\end{center}
\vspace*{5mm}
\caption{1-loop diagrams contributing to transmutation in Bhabha scattering.}
\label{bhabha1l}
\end{figure}
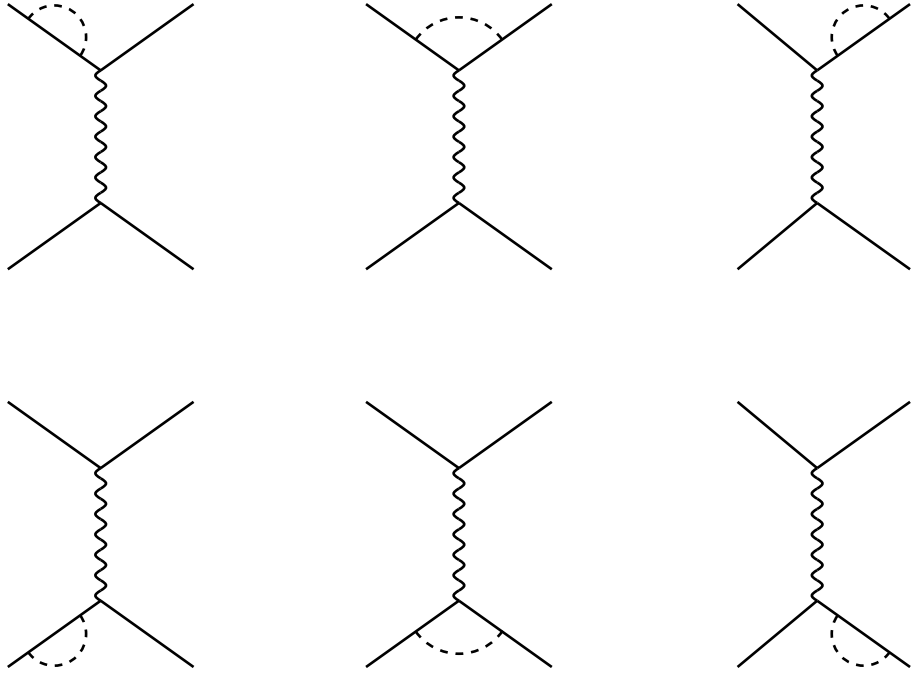

Fortunately for us, however, this estimate obtained from the kinematical
effects alone 
of the rotating mass matrix is not yet the full prediction of the 
DSM scheme.  In DSM, as detailed in section 3, the rotation of the fermion
mass matrix arises from insertions in the fermion propagator.  Thus, to
study transmutation effects properly to 1-loop order, one will
need to evaluate not just the 1-loop insertion in the fermion propagator
but all diagrams to the same 1-loop order.  For example, for the reactions
(\ref{comptone}) and (\ref{bhabha}), one will need to evaluate all the
diagrams in respectively Figure \ref{compton1l} and \ref{bhabha1l} plus
some others of no relevance to present considerations.  This calculation
has recently been done, and it was found that on summing all the relevant
1-loop diagrams, transmutation effects largely cancel leaving only terms
of order $s/M_X^2$ in amplitude \cite{transmudsm}, with $M_X$ being again
the generic mass of the mediating bosons carrying generation index.  In other
words, the net effect of transmutation, i.e.\  flavour-violation due to
mass matrix rotation, is just to give an additional contribution of the
same order as flavour-changing neutral current effects.

In DSM this cancellation is not an accident but is based on
quite general grounds.  Explicitly, the calculation goes as follows.  The
relevant 1-loop insertion to the fermion propagator (\ref{Sigma2})
can be rewritten in
the form:
\begin{equation}
\Sigma(p) = - \frac{\delta m}{\rho^2} + \frac{1}{2} (p\llap/ - m) B_L
   + \frac{1}{2} B_R (p\llap/ - m) + \Sigma_c(p),
\label{Sigma3}
\end{equation}
with
\begin{eqnarray}
B_L & = & - \frac{1}{16 \pi^2} \sum_K \int_0^1 dx (1 - x)
   \,\{ \bar{\gamma}_K^{\dagger}\,[\bar{C} - \ln (Q_0^2/\mu^2)]\,
   \bar{\gamma}_K \,\frac{1}{2}\,(1 + \gamma_5) \nonumber \\
   & & {} + \bar{\gamma}_K
  \,[\bar{C} - \ln (Q_0^2/\mu^2)]\, \bar{\gamma}_K^{\dagger}\,
   \frac{1}{2}\,(1 - \gamma_5) \}, \nonumber \\
B_R & = & - \frac{1}{16 \pi^2} \sum_K \int_0^1 dx (1 - x)
   \,\{ \bar{\gamma}_K \,[\bar{C} - \ln (Q_0^2/\mu^2)]\,
   \bar{\gamma}_K ^{\dagger} \,\frac{1}{2}\,(1 + \gamma_5) \nonumber \\
   & & {} + \bar{\gamma}_K^{\dagger} \,[\bar{C} - \ln (Q_0^2/\mu^2)]\,
   \bar{\gamma}_K
   \,\frac{1}{2}\,(1 - \gamma_5) \},
\label{BLBR}
\end{eqnarray}
and $\Sigma_c(p)$ of a form which need not bother us here except to note that
it is finite, independent of the renormalization scale, and of order $s/M_K^2$,
$M_K$ being, one recalls, the mass of the dual colour Higgs boson appearing
in the loop.  The insertion of (\ref{Sigma3}) to an internal fermion line
thus gives:
\begin{equation}
\frac{1}{p\llap/ - m} \longrightarrow \frac{1}{p\llap/ - m'}
   \,-\, \frac{\rho^2}{2} B_L \frac{1}{p\llap/ - m}
   \,-\, \frac{\rho^2}{2} \frac{1}{p\llap/ - m} B_R
   \,-\, \rho^2 \frac{1}{p\llap/ - m} \Sigma_c(p) \frac{1}{p\llap/ - m}
\label{internalp}
\end{equation}
and to an external fermion line:
\begin{eqnarray}
u(p) & \longrightarrow & u'(p) - \frac{\rho^2}{2} B_L u(p)
   - \rho^2 \frac{1}{p\llap/ - m} \Sigma_c(p) u(p) \\
\bar{u}(p) & \longrightarrow & \bar{u}'(p) - \frac{\rho^2}{2} \bar{u}(p) B_R
   - \rho^2 \bar{u}(p) \Sigma_c(p) \frac{1}{p\llap/ - m}
\label{externalp}
\end{eqnarray}
where $u'(p)$ is a solution of the Dirac equation with the renormalized
mass matrix $m'$:
\begin{equation}
(p\llap/ - m') u'(p) = 0.
\label{Diracp}
\end{equation}
These conclusions follow closely those in e.g.\ ordinary QED apart from
that, $m$ being a matrix and therefore noncommuting, $B_L$ and $B_R$ are
different.

A similar calculation gives as the vertex insertion:
\begin{equation}
\Lambda^\mu(p,p') = \frac{1}{2} \gamma^\mu L_L + \frac{1}{2} L_R \gamma^\mu
   + \Lambda_c^\mu(p,p'),
\label{Lambdamu3}
\end{equation}
with
\begin{eqnarray}
L_L & = & - \frac{1}{16 \pi^2} \sum_K \int_0^1 dx \int_0^x dy\; \{
   \bar{\gamma}_K^{\dagger} \,[\bar{C} - \ln (P^2/\mu^2)]\, \bar{\gamma}_K
   \,\frac{1}{2}\,(1 + \gamma_5) \nonumber \\
   & & {} + \bar{\gamma}_K \,[\bar{C} - \ln (P^2/\mu^2)]\,
   \bar{\gamma}_K^{\dagger}
   \,\frac{1}{2}\,(1 - \gamma_5) \}, \nonumber \\
L_R & = & - \frac{1}{16 \pi^2} \sum_K \int_0^1 dx \int_0^x dy\; \{
   \bar{\gamma}_K \,[\bar{C} - \ln (P^2/\mu^2)]\, \bar{\gamma}_K^{\dagger}
   \,\frac{1}{2}\,(1 + \gamma_5) \nonumber \\
   & & {} + \bar{\gamma}_K^{\dagger} \,[\bar{C} - \ln (P^2/\mu^2)]\,
   \bar{\gamma}_K
   \,\frac{1}{2}\,(1 - \gamma_5) \},
\label{LLLR}
\end{eqnarray}
where $\bar{C}$ is the divergent constant in (\ref{Cbar}),
\begin{equation}
P^2 = m^2(1 - y) + M_K^2 y - p^2 x(1 - x) - p'^2 (x - y)(1 - x + y)
   + 2pp'(1 - x)(x - y),
\label{Psquare}
\end{equation}
and $\Lambda_c^\mu(p, p')$ is finite, scale-independent and again of order 
$s/M_K^2$.

One notices that $L_L$ and $L_R$ in (\ref{LLLR}) are very similar in form
to respectively $B_L$ and $B_R$ in (\ref{BLBR}).  Indeed, it can be shown
that the pairs each differ only by terms of the order $s/M_K^2$.  This is
not really surprising, being just a consequence of gauge invariance, and
has a familiar parallel in ordinary electrodynamics.

With these results in hand, let us proceed now to calculate the amplitude
for $\gamma e$ collision to 1-loop order.  Adding to the tree diagrams the
1-loop diagrams of Figure \ref{compton1l}, and making use of the results
in (\ref{internalp}), (\ref{externalp}), and (\ref{Lambdamu3}), one obtains
the result:
\begin{eqnarray}
\hspace*{-10mm}& & \bar{u}'(p') \gamma^\mu \,\frac{i}{(p\llap/ + k\llap/) - m'}\,
   \gamma_\mu u'(p) \nonumber\\
\hspace*{-10mm}   & + 
& \frac{\rho^2}{2} \bar{u}(p') \,[\gamma^\mu (L_L - B_L) + (L_R - B_R)
   \gamma^\mu]\, \frac{i}{(p\llap/ + k\llap/) - m}\, \gamma_\mu u(p) 
\nonumber\\
\hspace*{-10mm}   & + & \frac{\rho^2}{2} \bar{u}(p') \gamma^\mu
    \,\frac{i}{(p\llap/ + k\llap/)
   - m}\, [\gamma_\mu (L_L - B_L) + (L_R - B_R) \gamma_\mu]\, u(p),
\label{ampegren}
\end{eqnarray}
plus only terms of order $s/M_K^2$ for large $M_K$.  We recall further that
in the differences $B_L - L_L$ and $B_R - L_R$, the divergent part and
the scale dependence both cancel, leaving in each only a finite part which
is again of order $s/M_K^2$ for large $M_K$.  Hence, for large $M_K$ the
renormalized amplitude (\ref{ampegren}) will reduce simply to the first
term there.  And this first term has no nondiagonal elements since by
definition, $u'(p')$ is a solution of the Dirac equation in (\ref{Diracp})
with mass $m'$ and therefore an eigenvector of the renormalized mass matrix
at any scale.  In other words, transmutation cancels here to order $s/M_K^2$
as claimed.  A similar analysis for $e^+ e^-$ leads to the same
conclusion.  Indeed, apart from some minor though nontrivial differences
these results closely resemble familiar results in ordinary electrodynamics.

The above result that transmutation effects or flavour-violations due to
mass matrix rotation cancel in DSM to order $s/M_X^2$, $M_X$ being 
the generic mass scale of flavour-changing neutral bosons, is a
great relief, since if they did not, they could easily lead to effects of
such a size as to contradict experiment, if not immediately then in the
very near future.  And as explained above, such a contradiction 
cannot easily be avoided by  readjusting parameters.  As
it is, the flavour-violating effects from mass matrix rotation is of the
same order as flavour-changing neutral current (FCNC) effects and can thus
be analysed together with the latter, to which we now turn.

Flavour-violation effects due to the direct exchange of flavour-changing
neutral or generation index-carrying bosons have so far been analysed by
us in DSM only for the gauge not the Higgs bosons, the reason being that
the former are to us theoretically better understood.  We believe, however,
that the results obtained would be similar for both.  The analysis follows
closely the usual pattern for other horizontal symmetry models, with
the predictions for flavour-violations strongly dependent on the exchanged
boson mass.  There is, however, one very important difference, namely that
the DSM scheme, being strongly constrained by what has gone before will now
be highly predictive.  Indeed, once given an estimate of the exchanged
gauge boson mass, then the DSM scheme so far developed will be able to
give quite precise predictions for the rates or cross sections of most
flavour-violation effects.  The reason is that if one accepts the tenets
of the DSM scheme, then both the coupling strength of the gauge boson and
its branching into various modes will be given.  The first will be given
by the Dirac quantization condition (\ref{Diraccond}) 
relating the required coupling ${\tilde g}$ of dual colour to the well-known
coupling of ordinary colour $g$, while the second will be given by the
orientations of the state vectors of the various fermion states relative both
to one another and to the gauge bosons, and these orientations are already
determined by the calculation in section 5 of the rotating mass matrix.  
For a first estimate of FCNC effects of relevance to the present experimental 
situation, only the 1-boson exchange diagrams need be calculated, higher 
order diagrams being suppressed by the high boson mass.  And these 1-boson 
exchange diagrams are calculable once given the gauge boson mass and the 
couplings to fermions.  Hence, apart from some technical details connected 
with the soft hadronic, nuclear or atomic physics inherent in respectively, 
for example, hadron decays, $\mu-e$ conversion in nuclei, and muonium 
conversion, which can be handled by almost standard methods, the calculations 
are relatively familiar and straightforward.  We need therefore only quote 
here some of the results as examples.

From an analysis of meson FCNC decays and mass splittings, one finds that
the strongest present bound on the flavour-violating gauge boson mass $M_X$
comes from $K_L - K_S$ mass splitting which gives a lower bound on the boson
mass of order $M_X/{\tilde g} \sim 400 \ {\rm TeV}$.  Taking this estimate
as benchmark value, and using the couplings to various channels deduced 
in the DSM scheme, one can then calculate the branching ratios for various FCNC
meson decays, some of the most interesting examples of which are listed in
Table \ref{fcncbr} \cite{fcnc}.  Further, applying the same considerations
to coherent $\mu-e$ conversions in nuclei with the same benchmark value
for the gauge boson mass, one obtains the conversion rates for some 
experimentally interesting nuclei as shown in Table \ref{muerates}
\cite{mueconv}.  One notices first that the predictions listed in both 
tables are quite detailed and precise for reasons already explained, and
secondly, that several of these are already close to the present experimental
limits.  This means that if any one of these FCNC effects is found in
experiment, hence giving an actual value for the gauge boson mass rather
than just a lower limit, then the correlated bounds listed in the 2 tables
can be used to give absolute predictions for all the others.  Or else, if
some other means is available to suggest a value for the gauge boson mass,
the same predictions can also be obtained.

\begin{table}
\begin{eqnarray*}
\begin{array}{||l|l|l||}
\hline \hline
  & {\rm Theory} & {\rm Experiment} \\
\hline
{\rm Br}(K^+ \rightarrow \pi^+ e^+ e^-)  &  4 \times 10^{-15}   &
2.9 \times 10^{-7}  \\
{\rm Br}(K^+ \rightarrow \pi^+ \mu^+ \mu^-)  & 2 \times 10^{-15}    &
7.6 \times 10^{-8}  \\
{\rm Br}(K^+ \rightarrow \pi^+ e^+ \mu^-)  &  2 \times 10^{-15}   &
7 \times 10^{-9}  \\
{\rm Br}(K^+ \rightarrow \pi^+ e^- \mu^+)  &  2 \times 10^{-15}  &
2.1 \times 10^{-10}   \\
{\rm Br}(K^+ \rightarrow \pi^+ \nu {\bar \nu})  &  2 \times 10^{-14}   &
1.5 \times 10^{-10}  \\
{\rm Br}(K_L \rightarrow e^+ e^-)  & 2 \times 10^{-13}   &
9 \times 10^{-12}  \\
{\rm Br}(K_L \rightarrow \mu^+ \mu^-)  & 7 \times 10^{-14}    &
7.2 \times 10^{-9} \\
{\rm Br}(K_L \rightarrow  e^{\pm} \mu^{\mp})  &  1 \times 10^{-13}    &
4.7 \times 10^{-12} \\
{\rm Br}(K_S \rightarrow \mu^+ \mu^-)  &  1 \times 10^{-16}   &
3.2 \times 10^{-7}\\
{\rm Br}(K_S \rightarrow  e^+ e^-)  &  3 \times 10^{-16}   &
1.4 \times 10^{-7} \\
\hline \hline
\end{array}
\end{eqnarray*}
\caption{Branching ratios for rare leptonic and semileptonic $K$ decays.
The first column shows the DSM predictions from one-dual gauge boson 
exchange with the boson mass scale at a benchmark value of 400 TeV.  The
second column gives either the present experimental limits on that process
if not yet observed or the actual measured value for that process.  In the
latter case, it means that the process can go by other mechanisms such as
second-order weak so that our predictions with dual gauge boson exchange
will appear as corrections to these.  All the empirical entries are from 
the data booklet \cite{databook}.}
\label{fcncbr}
\end{table}

\begin{table}
\begin{eqnarray*}
\begin{array}{||l|l|l||}
\hline \hline
{\rm Element}        & 
B_{\mu-e}^{\rm theor.} & B_{\mu-e}^{\rm exp.\ lim.} \\
\hline
^{27} {\rm Al}_{\,13}  & 1.4 \, 
\times \, 10^{-12} & {\rm n.a.}  \\
^{32} {\rm S}_{\,16}   & 1.9 \, 
\times \, 10^{-12} &
7 \, \times \, 10^{-11}  \\
^{48} {\rm Ti}_{\,22}  & 2.3 \, 
\times \, 10^{-12} &
 4.3 \, \times \, 10^{-12}\\
^{207} {\rm Pb}_{\,82} & 2.7 \, 
\times \, 10^{-12} &
4.6 \, \times \, 10^{-12} \\
\hline \hline
\end{array}
\end{eqnarray*}
\caption{Theoretical estimates for the ratio of the $\mu-e$ conversion rate
to the $\mu$ capture rate compared with present experimental limits.  These
values are calculated with the dual gauge boson mass scale taken at the
benchmark value of 400 TeV.}
\label{muerates}
\end{table}

Does there then exist any means for estimating the flavour-changing neutral
boson mass without first observing flavour-violation?  There is perhaps one
way which takes us interestingly to an entirely different area of physics
and another potentially dangerous prediction of the DSM scheme.  This
arises as follows.  The exchange of flavour-changing gauge bosons which
leads to flavour-violations is suppressed at low energies by the large
value of the boson mass to order $s/M_X^2$ in amplitude, which is the normal
reason given for FCNC effects being so small, and hence not yet
observed.  This is a copy of the explanation why ``weak'' interaction were
considered weak in the old days before the $W$ and $Z$ bosons were discovered
although the couplings of these bosons, as we now know, are by no means so
very small.  Indeed, it is by now a familiar fact that when experimental
energies rise beyond the $W$ and $Z$ mass, weak interaction cross sections
become sizeable.  In the same manner, therefore, for the processes mediated
by the even heavier flavour-changing bosons, cross sections will become
large also when energy is as large as the boson mass.  In fact, if one
believes DSM, then the effective interaction will become very strong given
that the boson coupling is constrained by the Dirac quantization condition
(\ref{Diraccond}) to be of order ${\tilde g} \sim 10$.  This means that
any particle carrying the generation index which allows it to couple
to the flavour-changing bosons and
hence interact by exchanging them will acquire very
strong interactions at high energies of order $\sqrt{s} \sim M_X$.  In
particular, even neutrinos which are believed now to exist in
generations are expected also to acquire this new interaction at high
energy.

This last is an astounding and at first sight very dangerous prediction,
for as we very well know, there is no indication at all in present day
experiment of such a phenomenon.  Indeed, we were very worried at first
until we realized that the gauge boson mass is constrained by present
bounds on FCNC effects, as explained above, to be larger than around 400
TeV, which is an enormous energy not achievable in the laboratory either
today or in the near future.  The only chance for observing effects at
such energy would be in cosmic rays, and even there the event rate would
be very, very small.  For a cosmic ray particle hitting a nucleon in the
atmosphere, $\sqrt{s} \sim 400$ TeV corresponds to a primary energy
of about $10^{20}$ eV.  Cosmic ray events at such primary energies
are known only in extensive air showers and even there are very rare,
incident on earth at an estimated rate of only about 1 event per square
kilometer per century.  Though rare, however, they have caught particular
attention, for ever since their observation in experiment they have been
a theoretical headache, for the following reason.

Although the origin of cosmic rays is still largely unknown so that primary
energies in excess of $10^{20}$ eV are in principle possible, there
is a bound, known as the GZK cut-off, on the energy of particles arriving
on earth from a distance of greater than about 50 Mpc.  Indeed, it was
shown by Greisen and by Zatsepin and Kuz'min \cite{Greisemin} already in
1966 that a proton or nucleus with such primary energies would quickly
degrade in energy by interacting with the 2.7 K microwave background via 
the following reaction:
\begin{equation}
p\,(N) + \gamma \longrightarrow p\,(N) + \pi,
\label{GZKreac}
\end{equation}
and hence not arrive on earth with their
primary energies intact, if they come from a distance of over 50 Mpc.  Yet,
over the years some 10 such events are claimed to have been seen, and they 
are beautiful things developing into a shower \cite{Baratov} with as many 
as $10^{11}$ charged particles!

So what are they?  According to Greisen, Zatsepin and Kuz'min, they cannot
be protons or nuclei coming from more than 50 Mpc away.  They are probably
also not from a nearby source, for within a radius of 50 Mpc, there are no
known astrophysical sources capable of producing particles with such high
energies.  Likewise, they are thought not to be photons, which can also 
interact with the microwave background by pair production and hence
cannot maintain their high energy over long distances.
And they cannot be neutrinos which can survive the journey but,
having only weak interactions, cannot produce air showers.  That is, unless
neutrinos can acquire strong interactions at these ultra high energies as 
predicted above by the DSM.

What would happen if one accepts that neutrinos do become strongly 
interacting for $\sqrt{s} \gtrsim 400$ TeV as the DSM suggests?  Then any 
source such as an active galactic nucleus which is capable of producing 
protons of such energies will also be able to produce neutrinos at these 
energies just by proton collisions via the said strong interactions.  Once 
produced, the neutrinos will be able to escape from the active 
galactic nucleus, although protons cannot because of the strong magnetic 
fields surrounding the source.  Further, the neutrinos will be able to 
survive the long journey through the microwave background in contrast to 
protons which cannot do so because of the GZK reaction (\ref{GZKreac}).  
And when the neutrinos arrive on earth, they will interact strongly by 
hypothesis with the air nuclei to produce the extensive air showers seen.  
Thus the hypothesis seems neatly to have passed all the initial tests and 
qualifies as a viable candidate solution to the GZK problem.  In addition,
it has even an explanation for a possible effect of these air showers 
reported by one experiment \cite{Agasa}.  Out of the dozen or so events
claimed to have been seen by this group, there are 3 doublets and 1 triplet
observed, the members of each multiplet being collimated within the 
experimental angular resolution of 1\,--\,2 degrees.  This suggests that 
members of each multiplet originate from the same source.  However, if 
the primaries were protons or nuclei, even when they originated from the same 
source, with in general different energies they would have been deflected 
differently by the intergalactic magnetic fields and 
lost their common direction.
Neutrinos, on the other hand, being neutral, would not be deflected by 
the magnetic fields and would remain collimated if they originate from 
the same source.

\begin{figure}[ht]
\centering
\includegraphics[angle=-90,scale=0.55]{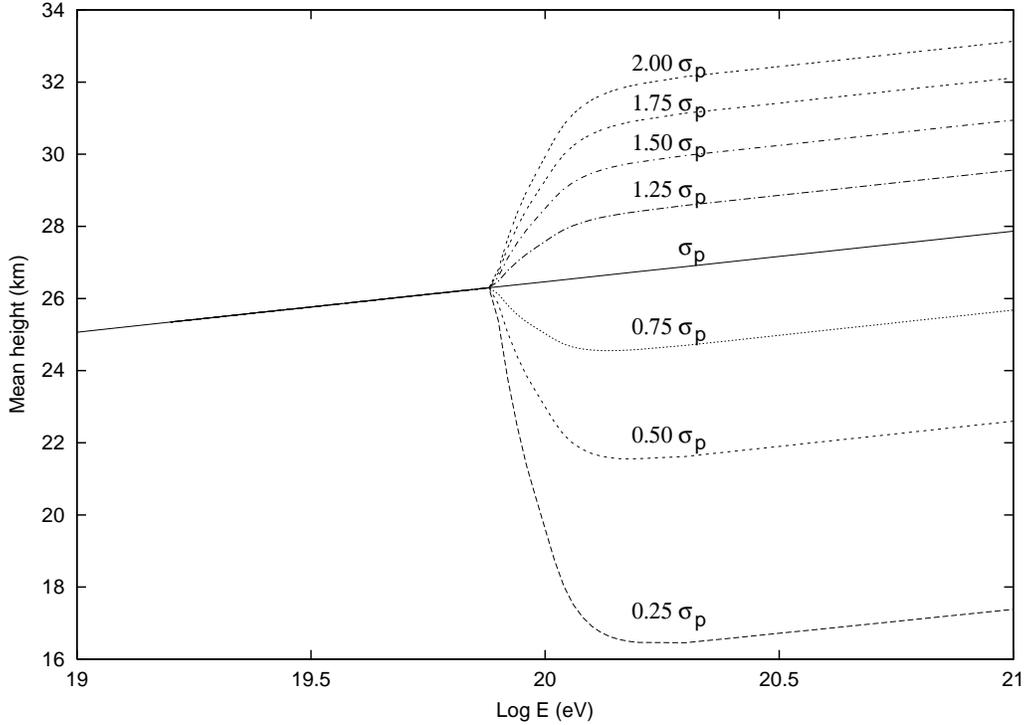}
\vspace*{5mm}
\caption{Average equivalent vertical height of air showers as a function
of the primary energy across the GZK cut-off assuming post-GZK primaries
of varying cross sections.}
\label{showerht}
\end{figure}

Interestingly, the above suggestion of post-GZK air showers being due to 
neutrinos acquiring strong interactions at high energy can be subjected 
to further experimental tests.  First of all, the hypothesis being that 
neutrinos interact strongly only at energies above the flavour-changing 
boson mass, it follows that the GZK threshold itself would put an upper 
bound on that mass.  Harking back then to our earlier discussion of FCNC
effects, this is exactly what is needed to constrain the magnitude of
these effects.  As it happens, the upper bound on the boson mass obtained
from the GZK threshold, as seen above, is close to the lower bound obtained 
before from $K_L - K_S$ mass splitting.  One would then obtain an actual 
estimate of the boson mass and could thus convert the previous bounds shown 
in e.g.\ Tables \ref{fcncbr} and \ref{muerates} into actual predictions.  
Although this estimate for the boson mass is very crude since all effects 
involved depend on the mass raised to the 4th power, nevertheless, it 
suggests in Tables \ref{fcncbr} and \ref{muerates} that FCNC effects may 
be just around the corner for experimental observation, and that these
predictions can soon be tested.  Secondly, in cosmic ray physics proper,
the hypothesis also suggests that air showers above the GZK cut-off should 
occur at lower heights in the atmosphere than those below the cut-off.  
The argument goes as follows.  The average height of air showers depend 
on the penetrating power of the incoming primary particle which in turn 
depends on the particle's cross section with air nuclei.  Now according 
to our hypothesis, pre-GZK showers are from protons or nuclei and post-GZK 
showers from neutrinos, and since protons and neutrinos presumably have 
different cross section with air nuclei, they should occur at different 
average heights.  In Figure \ref{showerht} is shown the change in average 
heights across the GZK threshold assuming different ratios of neutrino/proton 
cross sections with air nuclei.  In particular, if one assumes that the 
neutrino still appears as a point to the air nucleus at high energy, then 
the substitution of the known proton and air nucleus radii into a geometric 
picture gives an estimate for the ratio between neutrino and proton cross
sections of $\sigma_{\nu A}/\sigma_{p A} \sim 1/2$, corresponding to a 
change in height as shown in Figure \ref{showerht} which is sizeable.  
Such an effect may thus be looked for in Auger \cite{Auger} or other future
experiments.

Despite the scarcity and still 
somewhat tenuous nature of the data on post-GZK
air showers, we have described the DSM picture for them at some length,
first for the sheer beauty of these events and our own fascination with 
them, and secondly for the special role that they may play as a test for 
the DSM scheme.  As we have noted above already, despite the many tests 
to which the DSM predictions have been subjected and so far survived, not 
a single one hangs crucially upon the hypothesis that the horizontal 
symmetry of fermion generations is indeed identical to dual colour as 
suggested, 
apart from this one on post-GZK air showers.  
There was an important point in the above discussion that we have so far 
deliberately glossed over, namely that even given that neutrinos do acquire 
a strong interaction at extreme energies, it still does not mean that they 
will necessarily give air showers on collision with air nuclei, for which 
is needed not just a strong interaction between the neutrino and the air 
nucleus but a large hadronic-sized cross section.  An interaction between
the neutrino with the quarks inside the air nucleus which is
short-ranged,
as would happen if the horizontal symmetry has nothing to do with colour, 
say of the order of the mass $M_X$ of the heavy boson exchanged, will still 
give only minuscule cross sections no matter how strong it is.  However,
it was argued in \cite{airshow} that if the generation symmetry is indeed
identical to dual colour, then the neutrino at extreme energy will interact
not only strongly but coherently with the air nuclei, hence giving hadronic
sized cross sections sufficient to produce air showers in the atmosphere.
Indeed, this consideration is implicit in the geometric picture invoked
above to infer a $\nu A$ cross section of about a half of that of $p A$.
Although the argument given there is only qualitative, it is an important
point to bear in mind in considering the above explanation of post-GZK air
showers as initiated by strongly interacting neutrinos.

In summary, as far as flavour-violation and related questions are concerned,
which are the primary worry for the rotating mass matrix forming the basis 
of the DSM's main results on fermion mass and mixing patterns, the scheme 
seems to have survived all tests so far performed, and in the case of cosmic 
ray air showers it seems even to have offered a new explanation for an old
puzzle.  However, the job of surviving tests and limits is never done and
can be prolonged ad infinitum until one runs out of ideas or breath or
both.  We have performed more tests, which include for example an obvious
one on neutrinoless double-beta decay \cite{0nu2beta}, and which DSM also
survives.

\setcounter{equation}{0}

\section{Remarks}

In conclusion, it would seem that the DSM scheme has so far largely succeeded
in what it sets out to achieve, namely to suggest a {\it raison d'\^etre}
for 3 fermion generations and to explain their unusual mass and mixing
patterns.  Apart from CP-violation, even near quantitative results have
been obtained already with the 1-loop calculations performed in the energy
region where it is expected to be valid.  And in all areas explored up
to now where potential difficulties could arise, no violation of existing
experimental bounds is found.

There is, however, still one feature in the present formulation of the scheme
which leaves something to be desired.  The construction of the scheme as set 
out, for example, in Section 3, seems to come in 2 parts, first the theoretical
idea of deriving the generation symmetry and its breaking from duality, and 
second, the construction of a phenomenological model in terms of a Higgs 
potential and a Yukawa coupling, which though suggested by, cannot claim to 
follow from, the initial duality concepts.  And although duality does lead 
directly to the prediction of 3 and only 3 generations, the rest of the result
on mixing and so on
are consequences of the phenomenological model with at present
but tenuous links to duality.  Granted that even when considered as a purely 
phenomenological model, the derivation by itself of these latter results 
seems already not a mean achievement, one is nevertheless still some
way from being able to claim that the origin of generations as dual colour 
is now understood.

It seems to us, therefore, that to advance further, one should perhaps
proceed in 2 directions.  First, one should seek testable predictions of
the scheme which depend directly on the hypothesis that dual colour is
generation symmetry, 
of which we mentioned a possible example with post-GZK air
showers.  Secondly, one should try to derive directly from duality and
related concepts either the above phenomenological model itself or else
a scheme close to it which is capable of giving the same results.  This
is an ideal to strive for, but whether it can be achieved we do not know.

\newpage
\noindent{\large {\bf Acknowledgement}}

\vspace{.5cm}

Much of the original work which has gone into the above review, as well as
the manner in which
it is presented, has been achieved with the constant, 
close and always thoroughly enjoyable collaboration of Jos\'e Bordes, who 
only escapes being a co-author by the circumstantial accident of not being 
physically present at the Zakopane Summer School.


\begin{thebibliography}{99}

\bibitem{databook} Particle Physics Booklet, (2000), D.E.Groom et al.,
   extracted from the Review of Particle Physics, Euro. Phys. Journ. C15,
   (2000) 1.


\bibitem{Cabbibo}  N.\ Cabibbo, Phys.\ Rev.\ Lett.\ 10, 531 (1963).

\bibitem{CKM}  M.\ Kobayashi and T.\ Maskawa, Prog.\ Theor.\ Phys.\ 49, 652
   (1973).

\bibitem{MNS}  Z. Maki, M. Nakagawa and S. Sakata, Prog.\ Theor.\ Phys.
   28, 870 (1962).

\bibitem{SuperK}  Superkamiokande data, see e.g. talk by T. Toshito at
   ICHEP'00, Osaka (2000).

\bibitem{Soudan}  Soudan II data, see e.g. talk by G. Pearce, at ICHEP'00,
   Osaka (2000).

\bibitem{Sno} Q.R. Ahmad et al. Phys. Rev. Lett. 87, 071307, (2001),
   nucl-ex/0106015.

\bibitem{Homestake} B.T. Cleveland et al., Astropart. Phys. 496 (1998)
   505. 

\bibitem{Sage} J. Abdurashitov et al., Phys. Rev. C60 (1999) 055801.

\bibitem{Gallex} J.W. Hampel et al., Phys. Lett. B447 (1999) 127.

\bibitem{Chooz} CHOOZ collaboration, M. Apollonio et al., Phys.\ Lett.\
   B420, 397, (1998), hep-ex/9711002.

\bibitem{MSW}  L. Wolfenstein, Phys. Rev. D17, 2369, (1978); 
   S.P. Mikheyev and A.Yu. Smirnov, Nuovo Cim. 9C, 17 (1986).

\bibitem{Dirac} P.A.M. Dirac, Proc. Roy. Soc. London, A133, 60,
   (1931).

\bibitem{ourbook}  Chan Hong-Mo and Tsou Sheung Tsun, {\it Some Elementary
   Gauge Theory Concepts}, World Scientific, Singapore, 1993.

\bibitem{dualcomm}  Chan Hong-Mo and Tsou Sheung Tsun, Phys. Rev. D56,
   3646 (1997), hep-th/9702117.

\bibitem{WuYang}  Tai Tsun Wu and Chen Ning Yang, Phys. Rev. D14, 437 (1976).

\bibitem{Chanstsou2} Chan Hong-Mo, Peter Scharbach and Tsou Sheung Tsun,
   Ann. Phys. (NY) 167, 454 (1986).

\bibitem{Guyang} C.H. Gu and C.N. Yang, Sci. Sin. 28, 483 (1975).

\bibitem{Polyakov} A.M. Polyakov, Nucl. Phys. 164, 171 (1980).

\bibitem{Chanstsou1} Chan Hong-Mo, Peter Scharbach and Tsou Sheung Tsun,
   Ann. Phys. (NY) 166, 396 (1986); Chan Hong-Mo and Tsou Sheung Tsun,
   Acta Phys. Pol. B17, 259, (1986).

\bibitem{dualsymm} Chan Hong-Mo, Jacqueline Faridani, and Tsou Sheung Tsun,
   Phys. Rev. D53, 7293 (1996), hep-th/9512173.

\bibitem{horizontal} The idea of a possible ``horizontal symmetry'' linking
   generations is quite old.  Examples of some early references are:
   F. Wilczek and A. Zee, Phys. Rev. Lett. 42, 421, (1979);
   A. Davidson and K.C. Wali, Phys. Rev. D20, 1195, (1979), D21, 787, (1980);
   T. Maehara and T. Tanagida, Prog. Theor. Phys. 61, 1434, (1979);
   T. Yanagida, Phys. Rev. D22, 1826, (1980).

\bibitem{tHooft} G. 't~Hooft, Nucl. Phys. B138, 1 (1978); Acta Physica
   Austriaca Suppl. XXII, 531 (1980).

\bibitem{dualsym}  Chan Hong-Mo, Jacqueline Faridani, and Tsou Sheung Tsun,
   Phys. Rev. D52, 6134 (1995).

\bibitem{Heyl}  See, e.g., F.W. Heyl et al., Rev.\ Mod.\ Phys.\ 48,
   393, (1976).

\bibitem{physcons}  Chan Hong-Mo and Tsou Sheung Tsun, Phys. Rev. D57,
   2507, (1998), hep-th/9701120.

\bibitem{Weinberg} Steven Weinberg, Phys. Rev. D7, 2887, (1973).

\bibitem{ckm} Jos\'e Bordes, Chan Hong-Mo, Jacqueline Faridani, Jakov
   Pfaudler, and Tsou Sheung Tsun,  Phys. Rev. D58, 013004, (1998),
   hep-ph/9712276.

\bibitem{transmudsm} Jos\'e Bordes, Chan Hong-Mo and Tsou Sheung Tsun,
   Phys. Rev. D65, 093006, (2002), hep-ph/0111369. 

\bibitem{fcnc} Jos\'e Bordes, Chan Hong-Mo, Jacqueline Faridani, 
   Jakov Pfaudler,
   and Tsou Sheung Tsun, Phys. Rev. D60, 013005, (1999), hep-ph/9807277.

\bibitem{phenodsm} Jos\'e Bordes, Chan Hong-Mo and Tsou Sheung Tsun,
   Eur. Phys. J. C. 10, 63 (1999), hep-ph/9901440.

\bibitem{bordsm}  Jos\'e Bordes, Chan Hong-Mo and Tsou Sheung Tsun,
   hep-ph/0302199.

\bibitem{Ramon} H. Arason, D.J. Casta\~no, B. Kesthelyi, S. Mikaelian,
   E.J. Piard, P. Ramond, and B.D. Wright, Phys. Rev. D46, 3945, (1992).

\bibitem{Jarlskog} C. Jarlskog, in {\it CP Violation}, ed. C. Jarlskog,
   World Scientific, Singapore, 1989.

\bibitem{Docarmo} See e.g.\ L.P.\ Eisenhart, {\em A Treatise on the
   Differential Geometry of Curves and Surfaces}, Ginn and Company 1909,
   Boston; M.P.\ do Carmo, {\it Differential Geometry of Curves
   and Surfaces}, Prentice-Hall 1976, Englewood Cliffs, New Jersey.

\bibitem{features} Jos\'e Bordes, Chan Hong-Mo, Jakov Pfaudler, and
   Tsou Sheung Tsun, Phys. Rev. D58, 053006, (1998), hep-ph/9802436.

\bibitem{seesaw}  M. Gell-Mann, P. Ramond, and S. Slansky in
   {\it Supersymmetry}, ed. F. van Niuwenhuizen and D. Freeman
   (North Holland, Amsterdam, 1979); T. Tanagida, Prog. Theor. Phys.,
   B135, 66, (1978).

\bibitem{nuosc} Jos\'e Bordes, Chan Hong-Mo, Jakov Pfaudler, and
   Tsou Sheung Tsun, Phys. Rev. D58, 053003, (1998), hep-ph/9802420.

\bibitem{databook96}  R.\ M.\ Barnett et.\ al, Phys.\ Rev.\ D54, 1 (1996).

\bibitem{decoupth} T. Appelqvist and J. Carrazzone, Phys. Rev. D11
 (1975) 2865.  

\bibitem{cevidsm} Jos\'e Bordes, Chan Hong-Mo and Tsou Sheung Tsun,
   hep-ph/0203124, to appear in Eur.\ Phys.\ J.\ C.

\bibitem{K2K} S.H. Ahm et al., Phys. Lett. B511, 178, (2001), hep-ex/0103001.

\bibitem{transbhar} Jos\'e Bordes, Chan Hong-Mo and Tsou Sheung Tsun,
   hep-ph/0111175.

\bibitem{Babar} BaBar collaboration, see e.g. D.G. Hitlin, talk given at
   the 30th International Conference on High energy Physics, Osaka, Japan,
   Jul.-Aug. 2000, hep-ex/0011024.
\bibitem{Belle} Belle Collaboration, see e.g. paper presented at the
   7th International Conference on B Physics at Hadronic Machines,
   Kibutz Maagan, Israel, Sep. 2000, hep-ex/0101033.

\bibitem{mueconv} Jos\'e Bordes, Chan Hong-Mo, Ricardo Gallego, and
   Tsou Sheung Tsun, Phys. Rev. D61, 00077702, (2000), hep-ph/9909321.

\bibitem{Greisemin} K. Greisen, Phys. Rev. Letters, 16 (1966) 748;
   G.T. Zatsepin and V.A. Kuz'min, JETP Letters, 4 (1966) 78.

\bibitem{Baratov} See e.g. M. Boratov, in Proc. 7th International Workshop
   on Neutrino Telescopes, Venice (1996), astro-ph/9605087.

\bibitem{Takeda} See e.g. M. Takeda et al., Phys. Rev. Letters, 81, 1163,
   (1998) and references therein.

\bibitem{Agasa} M. Takeda et al., Astrophys. Journ. 522 (1999) 225;
   N. Hayashida et al., astro-ph/0008102.

\bibitem{Auger} The Pierre Auger Observatory Design Report (2nd ed.),
   14 March 1997.

\bibitem{airshow} J. Bordes, Chan Hong-Mo, J. Faridani, J. Pfaudler, and
   Tsou Sheung Tsun, Astroparticle Phys.\ 8, 135, (1998), astro-ph/9707031;
   see also hep-ph/9705463, (1997) unpublished; hep-ph/9711438, published in
   the {\em Proceedings of the International Workshop on Physics Beyond
   the Standard Model: from Theory to Experiment}, (ed.\ I Antoniadis,
   LE Iba\~{n}ez and JWF Valle) Valencia, October 1998, World Scientific
   (Singapore) 1998; Jos\'e Bordes, Chan Hong-Mo and Tsou Sheung Tsun,
   astro-ph/0012384.

\bibitem{0nu2beta} Jos\'e Bordes, Chan Hong-Mo, Ricardo Gallego, and
   Tsou Sheung Tsun, Phys. Rev. D63, 117701, (2001), hep-ph/0012242.


\end{thebibliography}
\end{document}